\documentstyle[12pt,epsfig]{article}
\newcommand{\beq}{\begin{equation}}
\newcommand{\eeq}{\end{equation}}
\newcommand{\bea}{\begin{eqnarray}}
\newcommand{\eea}{\end{eqnarray}}
\begin{document}
{\title{Hadron Diffractive Processes: the Structure of  Soft
Pomeron and  Colour Screening }
\vskip 1cm
\author{ L.G.Dakhno$^{1)}$ and V.A.Nikonov$^{2)}$}
\date{}
\maketitle

\vskip 1cm
\begin{center}
St.Petersburg Nuclear Physics Institute,\\
Gatchina, St.Petersburg 188350, Russia\\
$^{1)}$ dakhno@hep486.pnpi.spb.ru\\
$^{2)}$ nikon@rec03.pnpi.spb.ru
\end{center}

\begin{abstract}
On the basis of the experimental data on diffractive processes in
$\pi p$, $pp$ and $p \bar p$ collisions at intermediate, moderately
high and high energies, we restore the scattering amplitude related
to the $t$-channel exchange by vacuum quantum numbers by
taking account of the diffractive $s$-channel rescatterings.
 At intermediate and moderately high energies, the $t$-channel exchange
amplitude turns, with a good accuracy, into an effective pomeron which
renders the results of the additive quark model.  At superhigh
energies the scattering amplitude provides a Froissart-type behaviour,
with an asymptotic universality of  cross sections such as
$\sigma^{tot}_{\pi p}/\sigma^{tot}_{pp} \to 1$ at $s \to \infty$.
 The quark structure of hadrons being taken
into account at the level of constituent quarks, the cross
sections of pion and proton (antiproton) in the impact parameter
space of quarks, $\sigma_{\pi}(\vec {r}_{1\perp},\vec {r}_{2\perp};s)$
and $\sigma_p(\vec {r}_{1\perp},\vec {r}_{2\perp},\vec
{r}_{3\perp};s)$, are found as functions of $s$.
These cross sections implicate the phenomenon of colour screening: they
tend to zero at $|\vec r_{i\perp}-\vec r_{k\perp}| \to 0$.
The effective colour screening radius
for pion  (proton) is found for different $s$. The predictions for
the diffractive cross sections at superhigh energies are presented.

\end{abstract}
PACS: 14.20.Dh, 14.40.Aq, 13.85.Dz, 13.85.1g

\section{Introduction}
The investigation of diffractive processes at moderately high and
high energies turned rather long ago into the study of the
$t$-channel structure of the amplitude with vacuum quantum numbers
(pomeron).
The understanding of the $t$-channel exchange amplitude grew up
parallel with the growth of  energies  of the colliding particles
studied in the experiment. The main characteristics
  of the soft pomeron are its intercept and proper size. A
particular feature of the QCD pomeron is the colour screening
phenomenon.

In the latest decade it became clear that an understanding of the
$t$-channel structure of the amplitude is not enough for the
description of diffractive processes at high and superhigh
energies, because the $s$-channel diffraction rescatterings play here a
 crucial role.  The present paper is devoted to
the study of the soft pomeron (or Strong-QCD pomeron), with  a
simultaneous  $s$-channel unitarization of the
amplitude due to the account for diffractive rescatterings.

\vskip 1cm
 The paper is organized as follows.
In section 2 a brief survey of the problem is presented, with an
emphasis to various approaches to the Strong-QCD pomeron.
 In section 3, in the framework of generalized eikonal approach
which includes the $s$-channel diffractive rescatterings in the leading
order terms of the $1/N_c$-expansion,
the description of the experimental data is
carried out for the $p\bar p(pp)$, $\pi p$ and $\gamma p$ soft
diffractive processes.
In Conclusion we summarize the results underlining
our vision of diffractive processes.
In Appendix A the formulae for diffractive amplitudes are
derived, the basic points of the generalized eikonal
approach being  emphasized.
Appendix B is devoted to the three-reggeon amplitude {\bf PGG} which
is responsible for the colour screening. In Appendix C the
pomeron--meson amplitude is presented in terms of the light--cone
variables.

\section{Strong-QCD pomeron}
{\bf a. Pomeron intercept.}
The range of intermediate and moderately high energies is  well
described by the $t$-channel pomeron with an intercept
$\alpha_P(0)=1$ (the pomeron of Gribov - Chew - Frautschi (GCF)
\cite{Gribov_Chew}). Still,  with the energy increase it became
obvious that the experimental data do not obey  the GCF pomeron:
Kaidalov and Ter-Martirosyan (KT) \cite{KTM} suggested  a supercritical
 pomeron with an intercept $\alpha_P(0)=1+\Delta$
where $\Delta > 0$, that gives $\sigma^{tot} \sim s^{\Delta}$;
in fitting to data the magnitude $\Delta \simeq 0.12$ was found.
Donnachie and Landshoff (DL) \cite{DL} succeeded in the description of a
wide range of experimental data on the diffractive processes  using a
supercritical pomeron with $\Delta \simeq 0.08$.
 As was obvious from the very
beginning, the one-pomeron exchange amplitude with $\Delta \sim
0.1$ (KTDL pomeron)
is applicable in a restricted region of energies. With the
 energy growth, the one-pomeron exchange amplitude  with $\Delta \sim
0.1$ violates the unitarity requirement, for the maximal allowed growth
should be consistent with the Froissart limit which provides a
weaker growth, $\sigma^{tot} \sim \ln^2 s$.

In Refs. \cite{Sjost,Landshoff}, it was argued  that
 the pomeron with $\Delta =0.08$ is a universal object
capable to describe the soft diffraction  and small-$x$ deep inelastic
processes at
 moderately high and high energies. However, the value
$\Delta \simeq 0.08$ contradicts the  DESY
data on  vector-meson
electroproduction at the invariant energies $W=$10 - 200 GeV
\cite{el_photoprod}.  For this process  and
small-$x$ deep inelastic scattering
 the value $\Delta \sim 0.2 - 0.3$  was found
(see \cite{small-x} and references therein).

In the present paper we underline that the value $\Delta \sim 0.1$
does not provide a self-consistent description of  the diffractive
scattering data
even at moderately high energies.
The large diffractive cross section
(its magnitude is nearly one half of $\sigma^{tot}$, as was stressed
long ago \cite{ALR}) results in large $s$-channel--unitarity
corrections. In the complex-$j$ plane, they generate  branching points
of a considerable contribution, and these branching points are not
cancelled by other subprocesses. The total contribution which comes
from the pole and branching points requires  $\Delta > 0.1$
\cite{AD1,Anis_main}.

{\bf b. Soft-pomeron size.}
Pomeron is mainly a gluonic system, and this defines its properties.
One of the most important characteristics
of the pomeron, on which the physics of the
diffractive processes is founded, is the pomeron size. The
partons of the gluonic ladder,
which forms a pomeron, saturate a disk in the impact
parameter space. The radius of the disk
increases infinitely with  $s$, and
it is the pomeron slope, $\alpha'_P(0)$,
which determines the disk size at fixed
$s$.  As it follows from the experiment, the pomeron slope is not
large at modern high energies,
$\alpha'_P(0) \sim 0.2$ (GeV/c)$^{-2}$ \cite{alpha'}.
It can be
compared with the slope of the
$\rho$-meson trajectory -- this slope being a
typical hadronic value is considerably larger:
$\alpha'_{\rho}(0) \simeq 1$ (GeV/c)$^{-2}$.
For the pomeron represented by the gluon ladder of
Fig. 1a, the slope $\alpha'_P(0)$ is
determined by the size of the gluon plaquet which is a
constructive element for  this ladder, see Fig. 1b.
As was stressed in Ref. \cite{adnPR}, the small size of
this plaquet can be affected by a comparatively large mass of the
effective gluon (the soft gluon), this mass being of the order of
$0.7 - 1.0$ GeV.

The prompt evaluation of the effective gluon mass is possible in
radiative $J/\psi$ and $\Upsilon$ decays.
The estimation of the effective gluon mass
firstly performed by Parisi and Petronzio \cite{ParPet} for
 the decay $J/\psi \to \gamma+gg$ provided the value $m_g
\simeq 800$ MeV.  The analysis \cite{ConField} of more copious data
gave $m_g \simeq 800$ MeV for the reaction $J/\psi \to \gamma+gg$ and
$m_g=1100$ MeV for $\Upsilon \to \gamma+gg$.

The glueball physics   enlightens the problems of soft gluodynamics
as well.  Within the lattice calculus,
the following values were obtained for the mass
of the lightest scalar glueball: $m_{scalar \; glueball}=
 1549 \pm 53$ MeV
 \cite{ukqcd} and $1740\pm 71$ \cite{ibm}.
Experimental data also indicate the existence of the scalar-isoscalar
state in the mass region $1300 - 1600$ MeV:  this state
being an excess for the quark-antiquark nonet
systematics is a good candidate for the lightest scalar
glueball \cite{km1900,aasND}.  These results support a
comparatively large value of the effective gluon mass:
$m_g \sim m_{scalar \; glueball} /2 \sim  650 - 800$
MeV. Moreover,
 in line with this discussion,  lattice calculations  \cite{ibm}
resulted in a
small size of the lightest scalar gluodynamic glueball,
$<r^2>_{glueball} \simeq 0.1$ fm$^2$.

In the bootstrap-type model \cite{ags}, when the forces responsible for
the meson formation are determined by the exchange of both effective
gluon and mesons, the $q\bar q$-spectroscopic calculations of mesons
  from the low-lying multiplets $1^1S_0q \bar q$, $1^3S_1q \bar q$ and
 $1^3P_1q \bar q$ required the massive effective gluon, $m_g \simeq
  700$ MeV, in an agreement with the above-discussed values.

The  estimation of the gluon mass within perturbative QCD \cite{field}
 yielded the value of the same order:
  $m_g =1.5^{+1.2}_{-0.6}$ GeV, and recent lattice
calculation \cite{prd58} provided us with $m_g\simeq 0.8$ GeV.

The quark model had a striking success in the description of
hadron collisions at intermediate and moderately
high energies. The pomeron size is crucial characteristics for the
quark model, for only with a small-size pomeron  the additivity of
the scattering amplitudes occurs.
 This was  formulated rather long
ago \cite{anis_lec}. At 70's and the beginning of 80's, parallel with
accumulating experimental data, the arguments in favour of pomeron's
small size grew up \cite{anis_lec2}; they were summarized in
\cite{Book}. Later on, the small-size pomeron was named a point-like
pomeron \cite{Lands_Nacht}.

An opposite point of view concerning the pomeron had been developed
  in the approach initiated by Low and Nussinov \cite{LowNuss}, where
  the $t$-channel exchange with vacuum quantum numbers was treated
as an exchange of the two massless gluons (see Figs. 1c,d, where
typical diagrams for meson--meson scattering are drawn). The
$t$-channel massless gluons emulated a large radius for the $t$-channel
interaction, thus forming a large-size pomeron. A
noticeable advantage of this approach consisted in a
formulation of the  colour screening phenomenon
for  colliding quarks: the diagrams of Figs. 1c,d
cancelled each other at $|\vec
r_{1\perp}-\vec r_{2\perp}| \to 0$  ($\vec r_{1\perp}$ and $\vec
r_{2\perp}$ are impact parameters of quark and antiquark of a meson
represented by the upper block). Later on, the colour screening
phenomenon became a subject of a special discussion and
gave rise to the search for the colour transparency, see
\cite{Ralston} and references therein. In the Low--Nussinov model
the large radius of the two-gluon
interaction led to the dipole structure of the scattering
amplitude. As a result, $\sigma^{tot}_{\pi p}$  was
proportional to the pion mean radius squared, $<r^2_{\pi}>$, and
$\sigma^{tot}_{p p}$ was proportional to  $<r^2_{p}>$, hence
 $\sigma^{tot}(\pi p)/\sigma^{tot}(pp) \simeq
<r^2_{\pi}>/<r^2_{p}> \simeq 2/3$, in qualitative agreement with the
experimental data \cite{soper}.

However, one should stress that a direct use of the massless-gluon
exchange violated analytic properties of the
scattering amplitude:  the amplitude singularity appeared at
$t=0$ that made the discontinuity unequal to zero at $t>0$,
namely, $disc_t\, A^{el}_{\pi p,\,two-gluon\, exchange} \ne 0$ at
$0<t<4\mu^2_{\pi}$, while actually  $disc_t\, A^{el}_{\pi p}$ appears
only at $t>4\mu^2_{\pi}$.
  The region of small positive $t$ is in the
closest vicinity to the physical region of diffractive processes,
so the violation  of analyticity looks menacing here.
In \cite{NNNpot}, in order
to restore the  analyticity for the two-gluon exchange diagram,
a cutting was suggested,
 with $t_{cut}\sim 4\mu_{\pi}^2$; still, this parameter, as the
effective gluon mass estimation tells us, must be greater:
$t_{cut}\sim 2$ GeV$^2$.

{\bf c. Colour screening phenomenon.} The problem of taking
account of the colour screening for   quasi--point-like pomeron
 was discussed in \cite{adnPR},  where
 Lipatov's perturbative pomeron \cite{Lip} has been used as a guide.
The figure 1e to f demonstrates different couplings of the Lipatov-type
 pomeron to meson quarks: just
these two types of a coupling provide the
colour screening of meson quarks
for the soft pomeron. In further works \cite{LipRys,Muel_noimp}, the
problem of whether the colour screening was inherent to  Lipatov's
pomeron has been discussed in details.

According to
\cite{AD1,Anis_main}, the
corresponding meson--pomeron and proton--pomeron amplitudes
 have the following  structure:
$$
\rho(\vec {b}-\vec {r}_1)+\rho(\vec
{b}- \vec {r}_2)- 2\rho(\vec {b}- \frac{\vec {r}_1+\vec {r}_2}{2})
\exp(-\frac{(\vec {r}_1- \vec {r}_2)^2}{4r^2_{cs}})\,,
$$
\beq
\Sigma_{i=1,2,3}\;\rho(\vec {b}-\vec {r}_i)-
\Sigma_{i\ne k}
\;\rho(\vec {b}-\frac{\vec {r}_i+\vec {r}_k}{2})\exp(-\frac{(\vec {r}_i-
\vec {r}_k)^2}{4r^2_{cs}})\,.
\eeq
Here $\rho(\vec b)$ is the pomeron amplitude in the impact parameter
space, $\rho(\vec b-\vec r_{1\perp})$ corresponds to the pomeron-quark
vertex as is shown in Fig. 1e, where the pomeron interacts with one
quark only, the term $\rho(\vec b-\vec r_{2\perp})$ relates
to the second quark; the last term corresponds to the diagram of Fig. 1f
where the pomeron interacts with two quarks (see Appendices A, B, C for
details).
In the Lipatov-type pomeron, the $t$-channel gluons are reggeized
\cite{BFKL}, and
the intercept of reggeized gluon  is close to the unity, namely,
$\alpha_{gluon}(0)=1-\Delta_{gluon}$, where $\Delta_{gluon}$ is small.
The proximity $\alpha_{gluon}(0) \simeq 1$ is an essential point for
equation (1); we shall discuss it in Appendix B.

Actually, the equation (1) does not specify the pomeron size. A
large pomeron size responds to the case when one can neglect the
$r_{i\perp}$-dependence in the pomeron amplitude,
$\rho(\vec {b}-\vec {r}_i)\simeq \rho(\vec {b})$,
 while the exponent
is expanded in a series with respect to a small magnitude
$(\vec r_{1\perp}-\vec r_{2\perp})^2/4\vec r_{cs}^2$ (the value
$(\vec r_{1\perp}-\vec r_{2\perp})^2$ is restricted by  meson size,
and the colour screening parameter, $r_{cs}^2$, is assumed to be large).
The first non-vanishing term of a series
is proportional to $2(\vec r_{1\perp}-\vec r_{2\perp})^2$ for meson
and $(\vec r_{1\perp}-\vec r_{2\perp})^2+(\vec r_{2\perp}-\vec
r_{3\perp})^2+(\vec r_{1\perp}-\vec r_{3\perp})^2$
for proton that, after averaging over the meson (proton)
wave functions squared, results in the factor
$<r^2>_{meson}$ and $<r^2>_{proton}$,  in a complete similarity
to the two-gluon exchange model.  But from now on the similarity ends,
 for the amplitude (1) has as a factor the pomeron amplitude
$\rho(b)$, while for the two-gluon exchange model one has gluon
propagators.

For a small-size pomeron, the ratio $ r^2_{cs} /  <r^2>_{meson}$
is small, so
 the last term in  (1), which  implies the
colour screening, is small everywhere but for a region where
meson quarks are in a squeezed configuration:
$|\vec r_{1\perp}-\vec r_{2\perp}| \le r_{cs} $.
At the same time, the contribution of two first
terms, which correspond to the impulse approximation diagrams, is
significant everywhere where meson wave function dominates.
In such a way, the small-size pomeron justifies additive quark model
at moderately high energies; in particular, it leads to the ratio
$\sigma^{tot}(\pi p)/\sigma^{tot}(pp)
\simeq 2/3$ in accordance with the pion/proton constituent
 quark numbers.  Moreover, the  colour screening term in
 (1), though comparatively small,  allows one to
 explain the deviation from additivity within the  constituent quark
model at moderately high energies, that was a puzzle rather long ago.
 Namely, the ratio $\sigma^{tot}(\pi p)/\sigma^{tot}(pp)$ is
slightly less than 2/3 (experimentally it is $\frac{2}{3}(1-\delta)$,
with $\delta \simeq 0.1$). The Glauber rescatterings calculated
within constituent quark model (see Ref. \cite{Braun})
do not help providing $\delta \simeq
 -0.1$.  This is understandable, for the Glauber screening is
more significant for systems with larger number of constituents, that
is  for the proton. Therefore, the deviation of the cross
section ratio from 2/3, though small, is of a principal importance.
It was suggested in  \cite{anis_lec2} that just
three-reggeon diagrams with pomeron and two colour reggeons
are responsible for $\delta>0$. The
calculations of diffractive processes performed in
\cite{adnPR}  fortify this
idea specifying colour reggeons to be reggeized
gluons.

Using the language of reggeon exchanges, the impulse approximation diagrams
of the type of Fig. 1e can be represented as diagrams of Figs. 2a,b, so the
diagram of Fig. 1f which provides the colour screening is re-drawn as Fig.
2c; the shaded block of Fig. 2d represents the whole pomeron-meson vertex.
Here {\bf P} stands for the pomeron and {\bf G} for  reggeized gluon.  In
such a way the diagram responsible for the colour screening (Fig.  2c) is
the three-reggeon diagram {\bf PGG}.  Likewise, the  pomeron--nucleon
amplitude is shown in Fig.  2e to i, with all possible couplings of
reggeized gluons {\bf G} to nucleon quarks, thus granting the colour
screening.

{\bf d. The $s$-channel unitarity and soft primary pomeron.}
Within the one-pomeron exchange approach shown in Fig. 2
the calculations of the $pp$ and $\pi p$ scattering amplitudes
have been performed in \cite{adnPR}
 for $\sigma^{tot}$ and $d\sigma^{el}/dt$
in the energy range $p_{lab}\simeq 200\,-\,300$ GeV:
this is just the region where $\sigma^{tot}_{\pi p}$
and $\sigma^{tot}_{p p}$ are almost energy-independent, that corresponds
to $\alpha_P(0) \simeq 1$. Naively, the
problem of extending the model to the region of higher energies
looked  rather simple: one should
introduce $\Delta$ of the order of 0.1, following the suggestion of
 Refs.
\cite{KTM,DL}, and evaluate the corrections related to the two-pomeron
exchange (these corrections are to be taken into account, because
elastic and diffractive cross sections,
$\sigma^{el}$ and $\sigma^{DD}$, are determined by the imaginary parts
of the two-pomeron exchange diagrams,
and they are not small).  However, the realization of
this program with $\Delta \simeq 0.08$ faced a phenomenon which may be
called a hidden unitarity violation: the description of
experimental data with both one-pomeron (Fig. 3a) and
two-pomeron exchanges (diagrams of Fig. 3b type) requires $\Delta$
significantly larger than 0.08. The introduction of triple
rescatterings (Fig. 3c), etc. results in the subsequent increase of
$\Delta$; so one may conclude that the amplitude with $\Delta=0.08$ is
not self-consistently unitary to agree with the experiment even at
moderately high energies. The unitary amplitude which takes
into account a full set of the  $s$-channel rescattering diagrams (Fig.
3) and describes the available high-energy experimental data requires
\cite{Anis_main}:
\beq \Delta=0.29.
\eeq
We refer to the  soft
pomeron with $\Delta \simeq 0.3$ as a primary one.  A significant
increase of $\Delta$ as
a result of the $s$-channel rescatterings is due to
the large contribution of diffractive processes: even at moderately
high energies the diffractive processes provide nearly a half of
total cross section \cite{ALR}, and their rate approaches 1/2 with the
energy growth. A large rate of diffractive processes undermines the idea
thoroughly accepted at the early stage of the  pomeron study that
the $s$-channel unitarity of the scattering amplitude is mainly
due to the truly inelastic processes related to a pomeron
 cutting.

Fitting to experimental data on total and elastic cross
sections at high energies
\cite{Anis_main} provides us with the characteristics of
the primary pomeron strikingly close to the characteristics of
Lipatov's one \cite{Lip}.
Recall that the behaviour of amplitudes related to the
  diffractive hadronic processes is governed
by singularities  in the complex plane of the angular momentum $j$.
The pQCD pomeron  in the leading-logarithm approximation
  is a set of ladder diagrams with the
  reggeized $t$-channel gluons, see Fig. 1a
(the ladder diagram representation of the pQCD-pomeron
occurs with a special choice of the spin structure of the three--gluon
vertex, the gluons being reggeized). Within pure pQCD calculus,
the corresponding leading singularity in the
$j$-plane is the branching point at
$j=1+\Delta_{BFKL}$, where
$\Delta_{BFKL}=(g/\pi)^{2}3\ln2\simeq 0.5$ (BFKL-pomeron
\cite{BFKL,BFKL1}).
It is necessary to mention the latest
next-to-leading order ($\alpha_s^2$) calculations of $\Delta_{BFKL}$
\cite{FadKotLip} which provided the value $\Delta_{BFKL}\simeq
0.1-0.2$, though it might be probable that the next correction, of
the order of $\alpha_s^3$, would result in a value $\Delta_{BFKL}\simeq
0.3-0.4$ \cite{FadLip}.

The application of  QCD-pomeron to  phenomenological
calculations makes it urgent to consider the gluon virtual momenta
close to those used in the leading-logarithm approximation.  In
\cite{Lip} the virtualities of such a type have been effectively
taken into account with the help of a boundary condition, together with
the constraint ensured by the asymptotic freedom of QCD.  The pomeron
obtained in such a way, Lipatov's pomeron, is an infinite set of
poles in the region $1<j\le 1+\Delta$, and there exists a constraint
for the leading pole intercept:  $\Delta \ge 0.3$.

Coming back  to the results of \cite{Anis_main}, the vacuum
singularity of the primary pomeron  has
been approximated in the $j$-plane by the two poles:
\beq
j=1\;\;\; {\rm and} \;\;\;j=1+\Delta\;\;\; {\rm with}\;\;\; \Delta=0.29\,,
\eeq
following the idea \cite{NNNpot} of the two-pole
approximation of Lipatov's pomeron.

The value $\Delta \simeq 0.3$
obtained from the fit of total and elastic $\pi p$ and $pp$ processes
is on the lower boundary of the intercept of Lipatov's
pomeron. Moreover, the
small value of the
primary--pomeron slope, $\alpha'_P(0)$,
supports the
 idea about its small size. It was found \cite{Anis_main}
for the primary pomeron:
\beq
\alpha'_P=0.112\;\; ({\rm GeV/c})^{-2} .
\eeq
Another characteristics
of the primary pomeron, that is the colour screening radius
$r_{cs}$, also manifests the small size of the
primary  pomeron \cite{Anis_main}:
\beq
 r_{cs}=0.17 \;\; {\rm fm}.
\eeq
One may suggest that just the small size of the  primary
pomeron makes its properties close to the
characteristics to the Lipatov's pomeron.

In the present paper we have extended the applicability region,
having included the
intermediate energy region $p_{lab} \simeq 50-200$
GeV/c into consideration.  This requires an additional pole:
\beq
j=1-\Delta', \;\;\;{\rm with}\;\;\;\Delta'>0.
\eeq
However, this did not affect the characteristics of the
two leading poles  (3).

{\bf e. Superhigh energies and the Froissart limit.}
At superhigh energies, the diffractive cross section approaches its
asymptotics  determined by a full
set of multipomeron exchanges, which
examples are shown in  Fig. 3.
In the impact parameter space,  the interaction region is a black
disk, with the radius growing as $\ln s$ (Froissart limit).
At $\sqrt{s} \ge 500$ GeV, in the leading-logarithm
approximation (that means $R_{disk}\sim r_{disk} \ln s$), the
$p\bar p$ and $\pi p$ cross sections behave as follows \cite{Anis_main}:
$$\sigma^{tot}_{\pi p} \simeq \sigma^{tot}_{p\bar p} \simeq 2\pi
r^2_{disk}\ln^2 s\,,$$
$$\sigma^{el}_{\pi p} \simeq \sigma^{el}_{p\bar p} \simeq \pi
r^2_{disk}\ln^2 s\,,$$
\beq
\frac{d\sigma^{el}_{\pi p}}{dq^2_{\perp}} \simeq \frac{d\sigma^{el}
_{p\bar p}}{dq^2_{\perp}}
 \simeq \frac{\pi}{4} \,r^4_{disk}\,\ln^4 s
\, \exp\left (-\frac1{4}r^2_{disk}\ln^2 s\cdot q^2_{\perp}\right ),
\eeq
where $r^2_{disk}\simeq 0.051\; {\rm mb}=(0.071\;{\rm fm})^2$.
Generally, the scattering off the black disk leads to the Bessel--type
amplitude oscillations in $q^2_\perp$;
still, as we consider here the amplitude
in the region of small $q_\perp^2$, the exponential representation is
valid. It should be emphasized
that the late start of the asymptopia
is caused by small $r^2_{disk}$:
this magnitude almost  coincides with  $\alpha'_P=0.112\;({\rm
GeV/c})^{-2}=(0.066\;{\rm fm})^2$, see (4).  The dimensional
characteristics of the primary pomeron ($\alpha'_P,\;  r^2_{cs}, \;
r^2_{disk}$) obtained by fitting to data are rather small and of the
same order: we consider this as a manifestation of a comparatively
large mass of the effective gluon. The idea that the gluonic
structure of the high-energy $t$-channel exchanges results in a late
asymptopia was discussed in Ref. \cite{gl}.

 At lower energies, $\sqrt{s} \le 500$ GeV,
the terms proportional to $\ln s$  become important in diffractive cross
sections, and at  $\sqrt{s} <  15$ GeV
the term  $\sim 1/s^{\Delta'}$ is seen.

{\bf f. Intermediate and moderately high energies.}
The analysis of diffractive processes at
intermediate and moderately high energies, $p_{lab} \sim 50-100$ GeV/c,
 is of a great interest:\\
$(i)$ The pomeron exchanges dominanate at these energies,
so expanding our approach to this
energy region makes it possible to  resolve more precisely the
$j$-plane pomeron singularities.\\
$(ii)$ This energy region was discussed in the literature as a
 suitable one for the study of colour transparency. For
quantitative estimation of colour transparency, the colour
screening radius for the pomeron interaction should be known as a
function of energy: the analysis of diffractive processes can give such
information.\\
$(iii)$ The additive quark model provides rather good description of
data at intermediate and moderately high energies. So, it is very
instructive to trace the transitions of diffractive amplitudes
from superhigh energies, where hadron amplitudes are universal,
to the region, where the quark additivity works.

It should be stressed once again that coming from superhigh
energies, where pomeron parameters are properly determined, to the
region of moderate energies, where non-leading trajectories are
significant, allows us to evaluate more precisely the contributions of
the non-leading terms. This evaluation makes it possible to fix the
beginning of a pure pomeron contribution.

{\bf g. Eikonal and generalized eikonal approaches.}
The eikonal approach allows one to resolve the problem of the
$s$-channel unitarization of the diffractive amplitudes for $\Delta
>0$. The eikonal approach was applied to the $\pi p$ and $pp$
scattering amplitudes at high energies in
\cite{block,pumplin2}. However, the classic eikonal formulae do not
take into consideration the diffractive production processes in the
intermediate states, although this contribution is of the same order as
the classic eikonal rescattrings. In \cite{AD1,Anis_main} a generalized
eikonal approach  was developed in which the diffractive processes have
been considered  at the constituent quark level:
this allows us to take account of all diffractive
processes which are directly related to the disintegration of
colliding hadrons. This  is due to the quark--hadron duality:
 the $q\bar q$-state (see Fig. 4a) is equivalent to a full set of
hadron states (Fig.  4b,c).

The $1/N_c$ expansion rules \cite{1/N}
allow us to believe that generalized
eikonal is a reasonable approximation for the description of
diffractive processes caused by
primary pomeron. The fact is that the coupling of the primary pomeron
to hadron is suppressed as $1/N_c$, but this suppression is compensated
by the increase of $s^{\Delta}$. Multi-pomeron vertices, like {\bf PPP}
or {\bf PPPP} of Fig. 4d,e, are not included into generalized eikonal
approach because their contribution is even more suppressed:
 the vertex $g_{\bf PPP}$
is of the order of $1/N_c$ and  $g_{\bf PPPP} \sim 1/N_c^2$.

\section{Diffractive processes: the comparison\\ with  experiments}

In this section, in the framework of generalized eikonal approach,
we describe the elastic scattering amplitude for the $pp(\bar pp)$, $\pi
p$  and $\gamma p$
reactions. We consider $Im\, A^{el}(q^2=0) \sim \sigma^{tot}$,
$\sigma^{el}$, $\rho=Re\, A^{el}/Im\, A^{el}$
and the elastic scattering slope $B$ over wide energy
range, starting from $p_{lab}(pp)=50$ GeV/c, and restore the
characteristics of the primary pomeron.

The diffraction dissociation cross section
related to the dissiciation of the colliding hadrons
has been  calculated and compared
with the experimental data for $pp \to pX$, that allows us to estimate
the cross section which is due to the three--pomeron vertex {\bf PPP}.

{\bf a. Total and elastic cross sections.}  We refer to Appendices
A and B for the derivation of formulae written below. Our
 approach is applied to $\sigma^{tot,el}_{p\bar p(pp),\pi p}$,
the main goal is to extract the parameters of the soft
primary pomeron from the comparison with experimental data.
The following formulae
describe total and elastic  cross sections
of the colliding hadrons $A$ and $B$:
\beq
\sigma^{tot}_{AB}=2\int d^2b \int dr'\varphi^2_A(r')dr''\varphi^2_B(r'')
\left [1-\exp{(-\frac{1}{2}\chi_{AB}(r',r'',b)})\right],
\label{tot}
\eeq
\beq
\sigma^{el}_{AB}=\int d^2b
\left \{\int dr'\varphi^2_A(r')dr''\varphi^2_B(r'')
\left [1-\exp{(-\frac{1}{2}\chi_{AB}(r',r'',b)})\right ]\right\}^2\,.
\label{el}
\eeq
The expression  $dr\varphi^2_{A,B}(r)$ stands for  quark densities
of the colliding hadrons $A$ and $B$ which depend on the transverse
coordinates.  The pion and proton densities are defined as follows:
$$dr\,
\varphi^2_{\pi}(r)\equiv d^2r_1d^2r_2\,\delta ^2(\vec {r}_1+\vec {r}_2)
\,\varphi^2_{\pi}(r_1,r_2),
$$
\beq
dr \,\varphi^2_p(r)\equiv d^2r_1d^2r_2d^2r_3\,\delta ^2(\vec {r}_1+\vec
{r}_2 +\vec {r}_3) \,\varphi^2_p(r_1,r_2,r_3),
\eeq
where $r_i$ is transverse coordinate of the quark; the averaging over
longitudinal variables is performed.  The proton and pion quark
 densities are determined by their form factors; this is discussed in
 Appendix A.

The profile-function
$\chi_{AB}$ describes the interaction of quarks of colliding
hadrons via the pomeron exchange as follows:
\beq
\chi_{AB}(r',r'',b)=\int
db'db''\delta^2(b-b'+b'')\rho_A(b',r')\rho_B(b'',r'').
\eeq
The functions
$\rho_{A,B}$ stand for the amplitudes of the one--pomeron
exchange:
$$ \rho_{\pi}(\vec {r},\vec {b})=\rho(\vec {b}-\vec
{r}_1)+\rho(\vec {b}-\vec {r}_2)- 2\rho(\vec {b}-\frac{\vec {r}_1+\vec
{r}_2}{2})\exp(-\frac{(\vec {r}_1- \vec {r}_2)^2}{4r^2_{cs}}), $$
\beq
\rho_p(\vec {r},\vec {b})=
\Sigma_{i=1,2,3}\,\rho(\vec {b}-\vec {r}_i)-
\Sigma_{i\ne k}
\,\rho(\vec {b}-\frac{\vec {r}_i+\vec {r}_k}{2})\exp(-\frac{(\vec {r}_i-
\vec {r}_k)^2}{4r^2_{cs}}).
\label{ropirop}
\eeq
Functions $\rho_{\pi}$ and $\rho_p$
tend to zero at $|\vec {r}_{ij}| \to 0$.
We perform calculations in the centre-of-mass system of  colliding
hadrons, supposing that hadron momentum is equally shared between
quarks.  Then
\beq \rho(b)= \frac
{\sqrt{f_P(s_{qq})}}{4\pi(G+\frac{1}{2}\alpha'_P\ln{s_{qq}})} \exp\left
 [-\frac{b^2}{4(G+\frac{1}{2}\alpha'_P\ln{s_{qq}})}\right],
\label{rhoeff}
\eeq
where the pomeron-quark vertex $\sqrt{f_P(s_{qq})}$ depends on the
energy squared of the colliding quarks, $s_{qq}$ (see Appendix A for
details).

Equations (\ref{tot})--(\ref{el}) depend on the transverse
 coordinates of quarks only,
 though the
original expressions (C.9) and (C.17) of Appendix C
depend on the momentum fractions
 $x_i$ of  quarks of the colliding hadrons;
 hence $s_{qq}=sx_i^{(\pi)}x_j^{(p)}$ for $\pi p$ and
$s_{qq}=sx_i^{(p)}x_j^{(p)}$ for $pp$ collisions. We put $x_i=1/2$
for  meson and $x_i=1/3$ for  proton assuming
 that hadron wave functions $\varphi_{\pi}(\vec {r},x)$ and
$\varphi_p(\vec {r},x)$ select the mean values of $x_i$ in the interaction
blocks. A wide range of wave functions obey this assumption, for example,
the wave functions of the quark spectroscopy. But the
situation with  the colour screening diagram {\bf PGG} is more
complicated. One should integrate over a part of the
energy carried by reggeized gluons and pomeron: this spreads
$x_i$'s of interacting quarks. However, if the intercept
of the reggeized gluon $\alpha_G(0)\simeq 1$ that is actually a
requirement of the BFKL pomeron,  $x_i$ can be considered as
frozen. This assumption
is valid for $0.8<\alpha_G(0)<1$, that was
 checked by numerical calculations for realistic pion and proton wave
functions. In due course,  we put $s_{qq}=s/6$
for $\pi p$ and $s_{qq}=s/9$ for $pp$ collisions.

The equations (\ref{tot})--(\ref{el})
 can be used at small momentum transfers, when
real parts of the amplitude is small. We neglect the
signature factor
of the primary pomeron, though it can be easily restored.

The strategy for the total and elastic cross section calculations has
been chosen as follows. Initially we have included into our calculation
the region of high and superhigh
energies. The reason is that at such
energies the exchange of vacuum quantum numbers
(pomeron) is the only possible,
 the other contributions must vanish. Under such
an assumption, it is possible to restrict ourselves by the two leading
trajectory only. Therefore
\beq
\rho_{\pi}(\vec {r},\vec {b})=\rho^{(1)}_{\pi}(\vec {r},\vec {b})+
\rho^{(0)}_{\pi}(\vec {r},\vec {b})\,,
\eeq
$$
\rho_{p}(\vec {r},\vec {b})=\rho^{(1)}_{p}(\vec {r},\vec {b})+
\rho^{(0)}_{p}(\vec {r},\vec {b})\,.
$$
Every term $\rho^{(0,1)}(\vec {r},\vec {b})$ is given by
(\ref{ropirop}),
with its own set of parameters $G$, $\alpha'_P$, $r_{cs}^2$ and
$f_P(s_qq)$, but for high and superhigh energies the
identical sets of parameters for $\rho^{(0)}$ and $\rho^{(1)}$ occurred
to be a good approximation:
$G^{(1)}=G^{(0)}$, $r_{cs}^{(1)}=r_{cs}^{(0)}$,
$\alpha'^{(1)}_P=\alpha'^{(1)}_P$, and they differ by their
intercepts only.

In such a way the characteristics of the
primary pomeron have been found: the
 supercritical parameter $\Delta$, parameters $G$ and   $\alpha'_P$
for the pomeron slope, and colour screening radius:
\beq
\Delta=0.29,\;\;\;G= 0.167\; {\rm (GeV/c)}^{-2},\;\;\alpha'_P=0.112\;
{\rm (GeV/c)}^{-2},\;\;\;r_{cs}=0.17\;{\rm fm},
\label{param}
\eeq
and
\beq
f_P(s_{qq},q^2_\perp=0)
=g_1^2+g_0^2\,s_{qq}^{\Delta}\,,\;\;\Delta=0.29,\;\;
g_0^2=8.079\;{\rm mb},\;\;\;g_1^2=0.338\;\frac{{\rm mb}}{{\rm
GeV}^{2\Delta}}\,.
\label{pomhigh}
\eeq

For the description of intermediate energies it is necessary to
introduce at least
one more pole $\rho^{(-1)}(\vec {r},\vec {b})$, thus the whole
expression becomes:
\beq
\rho_{\pi}(\vec {r},\vec {b})= \rho^{(1)}_{\pi}(\vec {r},\vec {b})+
\rho^{(0)}_{\pi}(\vec {r},\vec {b})+
\rho^{(-1)}_{\pi}(\vec {r},\vec {b})\,,
\label{3_poles}
\eeq
$$
\rho_{p}(\vec {r},\vec {b})=\rho^{(1)}_{p}(\vec {r},\vec {b})+
\rho^{(0)}_{p}(\vec {r},\vec {b})+
\rho^{(-1)}_{p}(\vec {r},\vec {b})\,.
$$
In order to minimize the ambiguities, at $p_{lab} \sim 50-100$ GeV/c
the three terms in the r.h.s. of (\ref{3_poles}) are parameterized in
the form (\ref{rhoeff}); the parameters $f_P(s_{qq},q^2_\perp=0)$,
$\alpha'_P $, $G$ and $r_{cs}$ are considered as independent at
different energies. As it occurred, the parameter  $\alpha'_P$ remains
the same  as before: $\alpha'_P=0.112$ (GeV/c)$^{-2}$. The values
$f_P(s_{qq},q^2_\perp=0)$, $G$ and $r_{cs}$ are now
energy-dependent.  The function $f_P(s_{qq},q^2_\perp=0)$ is now as
follows:
\beq f_P(s_{qq},q^2_\perp=0)
=g_1^2\,s_{qq}^{\Delta}+g_0^2+\frac{g^2_{(-1)}}{s_{qq}^{\Delta'}}
\,,\;\;\Delta'=1.154\,,\;\;g_{(-1)}^2=-44.9\;{\rm mb}\cdot
{\rm GeV}^{-2\Delta'}\,.
\eeq

The description of total and elastic $p\bar p(pp)$ and
$\pi p$ cross sections at intermediate energies $\sqrt {s}=5 \div 20$
GeV is shown in Fig. 6a,b, together with the energy dependent
parameters (Fig. 6c,d,e). The accuracy for the parameter
$G$  is not high enough (Fig. 6e): its magnitude, like $\alpha'_P$, can
be regarded as energy-independent.  It should be pointed out that the
colour screening radius becomes smaller with the energy decrease; this
fact supports additive quark model at moderate energies (Fig. 6d).
A later (in the $s$-scale) onset of three-regeon diagram {\bf
GGP} compared with the onset of one-pomeron diagram {\bf P} results
in a much smaller colour screening effects at $\sqrt {s} \sim 5 $ GeV
(in {\bf GGP} the energy  $\sqrt{s}$ is shared between {\bf G} and {\bf
P}:  $\sqrt{s}m^2 =\sqrt{s_{\bf G}s_{\bf P}}$ where $m \sim 1$ GeV).
In other words, the colour screening is somehow analogous to the
inelastic shadowing which starts at the energies large enough to
exceed the inelasticity threshold in the intermediate state. Then, with
a further increase of energies, the contribution of inelastic shadowing
grows, until the integration over intermediate mass is saturated by the
three--pomeron peak, or --- as it happens with colour screening --- by
{\bf PGG} peak, which behaves like {\bf PPP} due to the proximity of
the reggeized gluon intercept to the unity. Because of this similarity,
it looks natural that colour screening radius $r_{cs}$ tends to zero
with the energy decrease.

At asymptotic energies ($\sqrt{s}\ge 500$ GeV)
the cross sections $\sigma^{tot}_{pp}$ and
$\sigma^{tot}_{\pi p}$ calculated with the parameters (\ref{param})
increase as $0.32\ln^2s$ mb, while the growth with energy of the
elastic cross sections, $\sigma^{el}_{pp}$
and $\sigma^{el}_{\pi p}$,  is proportional to $0.16\ln^2s$ mb.
At $\sqrt{s}\ge 50$ GeV, within the 5\% accuracy,
the  calculated total
and elastic cross sections can be approximated
by the following formulae:
$$\sigma^{tot}_{pp}=49.80+8.16\ln (s_{qq}/s_0)+0.32\ln ^2 (s_{qq}/s_0),
$$
$$\sigma^{tot}_{\pi p}=30.31+5.70\ln (s_{qq}/s_0)+0.32\ln ^2
(s_{qq}/s_0).$$
$$
\sigma^{el}_{pp}=8.19+3.027\ln (s_{qq}/s_0)+0.16\ln ^2 (s_{qq}/s_0),
$$
\beq
\sigma^{el}_{\pi p}=3.87+1.567\ln (s_{qq}/s_0)+0.16\ln ^2 (s_{qq}/s_0).
\label{asym}
\eeq

In (\ref{asym}) the numerical coefficients
are given in mb and $s_0=10000$
GeV$^2$.  Recall that $s_{qq}=s/9$ for $pp(p\bar p)$ and
$s_{qq}=s/6$ for $\pi p$ collisions.

Our predictions for LHC energies ($\sqrt{s}=16$ TeV)  are:
$$\sigma^{tot}_{p\bar p}=131\;\;\; {\rm mb},\qquad
\sigma^{el}_{p\bar p}=41 \;\;\; {\rm mb}.$$

At far asymptotic energies
the ratio of  total cross sections
$\sigma^{tot}_{pp}/\sigma^{tot}_{\pi p}$ tends to the unity.
At these energies $\sigma^{el}_{AB}/\sigma^{tot}_{AB} \to 1/2$
(black disk limit).

{\bf b. Diffraction dissociation of colliding hadrons.}
The following formulae stand for the
diffraction dissociation processes:
\beq
\sigma^{single,DD(B)}_{AB}
+\sigma^{el}_{AB}=
\int d^2 b \int dr'\varphi^2_A(r')dr''\varphi^2_B(r'')
d\tilde{r}'\varphi^2_A(\tilde{r}')
\label{single}
\eeq
$$\times\left [1-\exp{(-\frac{1}{2}\chi_{AB}(r',r'',b)})\right]
\left [1-\exp{(-\frac{1}{2}\chi_{AB}(\tilde{r}',r'',b)})\right],$$
\beq
\sigma^{total \;diffraction}_{AB}=
\sigma^{el}_{AB}+\sigma^{single,DD(B)}_{AB}+\sigma^{single,DD(A)}_{AB}
+\sigma^{double}_{AB}
\label{double}
\eeq
$$=\int d^2b
\int dr'\varphi^2_A(r')dr''\varphi^2_B(r'')
\left [1-\exp(-\frac{1}{2}\chi_{AB}(r',r'',b))\right]^2\,,$$
where $\sigma^{single,DD(B)}_{AB}$
describes the  diffractive dissociation of a hadron B and
$\sigma^{total\;diffraction}_{AB}$, stands for total hadron
diffraction.

Let it be emphasized that there are two mechanisms
contributing  to the
diffractive dissociation cross section
$\sigma^{single,DD(B)}_{AB}$
measured at the experiment:\\
$(i)$ dissociation of a colliding
hadron, see Fig.  7a, and \\
$(ii)$ partly dissociated pomeron, Fig.  7b (the cross section for the
process of Fig. 7b is shown separately in Fig. 7c: it is related to the
three--pomeron cut).

In the used approach
the formulae  (\ref{single})--(\ref{double})  describe the hadron
dissociation only but not the  pomeron one.
The calculated cross section $\sigma^{single,DD(p)}_{pp}$ which
is due to the proton dissociation
is presented in Fig. 7d,
and figure 7e shows  the difference
$\sigma^{single,DD(p)}_{pp}(exp)-
\sigma^{single,DD(p)}_{pp}(calculated)$,
the latter term given by Eq. (\ref{single}). This difference stands
just for the diffraction  of a partly dissociated pomeron, that is the
 three--pomeron diagram of Fig. 7c.  It should be pointed out that
generalized eikonal
  approach allows one to calculate the other characteristics
of the diffractive hadron dissociation, namely, the $M^2$- and
$t$-dependences.  However, such a study  is beyond the scope of this
article.

{\bf c. Photon--proton total cross section and the
photo-production of vector mesons.}
The developed approach, which until now has been applied to
the diffractive  $pp$ and $\pi p$ cross sections,
can be also applied to the reactions with a photon, that is based
 on the hypothesis of vector-meson dominance, $\gamma \to V$.
 Corresponding calculations have been
performed for the sum of diagrams shown in Fig. 8a,b. It is assumed that
the wave functions of  vector mesons ($\rho$, $\omega$, $\phi$) are
equal to that of pion's, $\psi_V \simeq \psi_{\pi}$, as they are the
members of the same SU(6)-multiplet.

The total cross section $\sigma^{tot}_{\gamma p} $
has been calculated, with the same parameters
for the  primary pomeron
which have been found for the reactions $pp$ and $\pi p$. The extra
constant is the normalization parameter which
determines the transition $\gamma \to V$; its value is defined by
$\sigma^{tot}(\gamma p)$ at $\sqrt{s}=W_{\gamma p}=20$ GeV. The
results of the calculation are shown in Fig. 8c, together with the
available experimental data (see \cite{ZEUS_H1} and references
therein).

  In Fig. 8d we demonstrate the cross section
$\gamma p \to \rho/\omega \;p$
calculated under the assumption
$\sigma(\gamma p \to \rho p)=\sigma(\gamma p \to \omega p)$. No new
parameter is used comparing with the calculation of
$\sigma^{tot}_{\gamma p} $.

{\bf d. Effective colour screening radius $r_{cs}^{eff}$.}
The concept of colour screening which is realized here on the basis of
gluon structure of the pomeron makes it necessary to introduce, apart
from the colour screening radius of a primary pomeron, the effective
colour screening radius.
 For the pion--proton interaction, the colour screening profile factor
is determined as:
\beq
\zeta_{\pi}(r,s)=N
\sigma_{\pi}(\vec {r}_{1\perp},\vec {r}_{2\perp};s)=N
\,\int d^2b\, dr''\,\varphi^2_p(r'')
\left [1-\exp\left (-\frac12\chi_{\pi p}(r,r'',b)\right )\right]\,.
\eeq
Here $N$ is a normalization factor which is chosen to satisfy the
constraint:
$$\zeta_{\pi}(r \to \infty,s)=1.$$
The physical meaning of this colour screening profile factor
 is simple: the pion-proton interaction $\zeta_{\pi}(r,s)$
depends on the pion
interquark distance, and it tends to zero with $r\to 0$, in line with
general concept of the colour screening.
After the integrations over  the impact parameter $b$ and      proton
coordinates $r''$, we have found
 $\zeta_{\pi}(r)$ which is shown in Fig. 9 for
different $s$.
 At     far
asymptotic energies $\sqrt{s} >> 10^{10}$ GeV$^2$
 the effective colour screening radius tends to zero. This
phenomenon comes due to the diffusion of the pomeron gluons
in the impact parameter space.

For moderate and high energies a simple approximate
function is useful:
\beq
\zeta_{\pi}(r,\sqrt{s} \sim 25-1800
\;{\rm GeV})=1-\exp\left[-(r/r_{cs}^{eff})^n\right]\,,
\eeq
with  $n=1.89$ and $r_{cs}^{eff}=0.172$ fm.

Likewise, effective colour screening radius has been defined for
the proton; its numerical value practically coincides with that of
pion's. It is worth noting that the effective radius $r_{cs}^{eff}$
and colour
screening radius of the primary pomeron $r_{cs}$
are very close to each other at
$\sqrt{s} \sim 25-1800$ GeV.

{\bf e. The low-energy effective pomeron.}
The last point to be discussed in this section is how the performed
calculations relate to the description of the diffractive
processes at intermediate and moderately high energies.
 In the paper \cite{adnPR}  the $\pi p$ and
$pp$ diffractive cross sections were simultaneously described at FNAL
energies,  in the framework of the one-pomeron exchange with
colour screening taken into account.
 The $qq$-amplitude
 calculated in \cite{adnPR} has been found to be equal to 5.5 mb.
Let us consider this single pomeron  as an effective
one {\bf P}$_{eff}$ and compare it with analogous magnitude obtained
within generalized eikonal approach developed here. The effective
pomeron is actually a sum of  multi-pomeron exchanges  of a primary
pomeron shown in Fig. 10a,b, the colour screening neglected.
The summation of all the pomeron graphs of Fig. 10a,b provides the
value 6 mb for $f^{(P_{eff})}_{qq}=\sigma^{tot}_{qq}$ at $\sqrt{s}=24$
GeV, thus revealing a self-consistency of both approaches.
It should be noted that $\sigma^{tot}_{qq}=f_P(s_{qq},q^2_\perp=0)$
of the primary pomeron is 9 mb at this energy and falls down to 6 mb
at $s_{qq}$=10 GeV$^2$ (see Fig. 6c). This means that multiple
rescatterings are not much significant, justifying the results of the
additive quark model in this region.

Calculated at different energies, this quark-pomeron amplitude
$f^{(P_{eff})}_{qq}(s,q^2_\perp=0)$ is shown in Fig. 10c; it should be
emphasized that asymptotically this magnitude,
$f^{(P_{eff})}_{qq}(s,q^2_\perp=0)$, increases as $\ln^2 s$.

\section{Conclusion}
In this article we have performed the description of soft diffractive
processes in the $pp(p\bar p)$, $\pi p$ and $\gamma p$ processes
within the
framework of generalized eikonal approximation at the whole range of
available energies, the characteristics of a soft primary pomeron
have been found. Generalized eikonal approximation is a correct
representation of the $s$-channel unitarized amplitude with respect to
the leading-in-$s$ terms, provided the multi-pomeron vertices
({\bf PPP}, {\bf PPPP}, etc) are suppressed in the $1/N_c$
expansion ($g_{{\bf PPP}} \sim 1/N_c$, $g_{{\bf PPPP}} \sim 1/N^2_c$,
etc).

The characteristics of the soft primary pomeron occurred to be in a
 proximity to those of the pQCD pomeron
(Lipatov's pomeron \cite{Lip}). The primary
pomeron is approximated by three poles in the complex plane $j$:
$$
j=1-\Delta',\;\;\;1,\;\;\;1+\Delta, \;\;\;{\rm with}\;\;\;\Delta=0.29
\;\;\;{\rm and}\;\;\; \Delta'=1.154\,.
$$
The intercept of the leading pole is close to that of leading pole of
Lipatov's pomeron. A small proper size of the soft primary pomeron,
by our opinion, causes this proximity.

Mainly, the asymptotic behaviour of cross sections is
determined by the leading pole
 with multiple rescatterings, that leads to
a Froissart--type growth $\sigma^{tot} \simeq 2\pi r^2_{disc}\ln^2 s$.
The coefficient $r^2_{disc}$ is of the
order of $\alpha'_{\bf P}(0)$: $r^2_{disc} \simeq  \alpha'_{\bf P}(0)$.
A quasi point-like structure of the
primary pomeron is directly related to the
small value of parameters $r^2_{disc} = 0.051$ mb
$= (0.071$ fm$)^2$ and
$r_{cs}=0.17$  fm, that is unambiguously connected with the
large effective soft gluon mass defined in the analysis of data on
radiative $J/\psi$ decay.

At asymptotic energies $r^{eff}_{cs}(s)$ tends to 0 that is due to
a large diffusion of partons in the pomeron ladder.
 With the energy decrease, the effective colour screening radius
grows, being of the order of the primary pomeron radius in the interval
$\sqrt{s} \sim 50-10^{10}$ GeV. The further descent to the intermediate
energies, such as $p_{lab} \sim 50-100$ GeV/c, results in a
decrease of the colour screening radius of the primary pomeron, that
makes the effective radius much smaller too, justifying additive quark
model.

The scattering amplitude which is obtained within the framework of
generalized eikonal approach in the energy range $\sqrt{s}=25-1800$ GeV
increases weakly, like $ s^{0.1}$. This is just the region where the
amplitude reproduces the behaviour of the KTDL-pole. The growth
of $\sigma^{tot}_{\pi p}$, $\sigma^{tot}_{p p}$
and $\sigma^{tot}_{\gamma p}$ is found to be universal that does not
agree with the statement of \cite{DonLan} that shadowing results in
process-dependent apparent intercepts for these reactions.

At intermediate energies, $p_{lab} \simeq 50-100$ GeV/c, all three
 poles $j<1$ and $j=1$ are significant.
In this region the
generalized eikonal approach reproduces qualitatively the results of
the quark model. The colour
screening radius decreases significantly, $r^2_{cs}(s_{qq})\sim
5\;\;{\rm GeV}^2)=0.02\pm 0.01$ fm$^2$, thus reducing the
colour screening effects into amplitude. This is quite natural for
intermediate energies because the colour screening, in terms
of hadronic language, is due to inelastic shadowing
which is small; it increases steadily with the energy growth and is
stabilizing at high energies (see, for example, \cite{Ralston,Giann}
and references therein).

The performed analysis allows us to conclude that primary pomeron,
which properties are close to those of Lipatov's pomeron, is a
universal object for the description of soft diffractive processes in
the whole interval of high energies, starting from $\sqrt{s}\sim 25$
GeV.  We would like to stress  that primary pomeron with
$\Delta\simeq 0.29$, colour screening and multiple rescatterings
included, describes simultaneously the data on $\pi p$, $pp$ and
$\gamma p\to Vp$ reaction, the growth rate being nearly the same. This
fact does not agree with the statement made in \cite{DonLan} that
colour screening affects the different growth rates of cross sections
for different reactions.

Concluding, we would like to underline the basic difference of the
developed approach from that of Donachie--Landshoff
\cite{DL,Landshoff,DonLan}. In \cite{DL,Landshoff} soft
diffractive amplitudes are due to the soft pomeron exchange with
$\Delta_{soft}=0.08$, while a new object -- hard pomeron with
$\Delta_{hard}=0.3$ -- is introduced for the vector meson
electroproduction processes $\gamma*(Q^2)V \to VP$
\cite{DonLan}. The hard pomeron
vertex $\gamma^*(Q^2)\,V \to hard\;pomeron$ depends on $Q^2$, being
rather small or equal to zero  at  $Q^2=0$. However, one may expect
that a realization of this hypothesis in terms of quarks or
hadrons needs a special dynamics: the problem is how to relate
this dynamics to the vector dominance idea or, more generally,
to the mechanism of the photon hadronization $\gamma \to q\bar q$.

In our model it is the primary pomeron  who has $\Delta\simeq
0.29$, and a weak cross section growth at $\sqrt{s}=50-2000$ GeV is due
to a considerable shadowing which appeared to be universal for the
light hadrons and photon (within photon hadronization).
 One may believe that in the reaction
 $\gamma^*(Q^2)V \to VP$ the shadowing effects should vanish at large
$Q^2$, leaving the one--pomeron exchange responsible for the process at
$Q^2 \sim 10-20$ GeV$^2$.

\section*{Acknowledgement}
The authors are grateful to V.V. Anisovich, Ya.I. Azimov, L.N. Lipatov and
M.G.  Ryskin for useful discussions and comments. V.A.N. acknowledges
the support of the RFBR grant N 98-02-17236.

\section*{Appendix A. Soft pomeron and the s-channel unitarized
amplitude: meson-meson elastic scattering }

Here we present the formulae for the  amplitudes of
diffractive processes emphasizing the basic points of our
approach. To illustrate the method, we consider as an example  the
meson-meson scattering amplitude that allows us to underline
general features of the $s$-channel unitarization procedure.
Then the formulae for the
pion-nucleon and nucleon-nucleon (or nucleon-antinucleon) diffractive
scattering are presented; some of them were given in
\cite{Anis_main}, though without derivation.

The study of the meson-meson scattering amplitude
(as well as the other diffractive amplitudes) is performed
in the impact parameter space, that is suitable
for the
$s$-channel unitarization .  The consideration
is carried out in the following way:\\
$(i)$ First, we consider the
impulse approximation  diagram for the exchange of the primary pomeron
{\bf P} -- the interaction of the type of Fig. 2a-b for meson-
or Fig. 2e-g for proton-pomeron amplitude.
For the exchange of primary pomeron,
 the standard eikonal unitarization is performed.   \\
$(ii)$ Then, the three-reggeon {\bf PGG} and
five-reggeon  {\bf GGPGG} diagrams are considered: these diagrams
are responsible for the colour screening in
the primary pomeron exchange amplitude.     \\
$(iii)$ As a last step, we take into account
a full set of primary pomeron interactions
({\bf P}, {\bf PGG}, {\bf GGP} and {\bf GGPGG}) in the
generalized eikonal approximation (Fig. 3):
it provides a unitarized  meson-\-meson scattering
amplitude with a colour screening.

{\bf a. Primary pomeron exchange and the eikonal unitarization.}
Within the standard normalization, the soft-scattering
amplitude of Fig. 11a reads:
$$
A^{( P)}(s,q^2_{\perp}) =
i\,s\,F_A(q^2_{\perp})F_B(q^2_{\perp})P(s,q^2_{\perp}).
\eqno{(A.1)}
$$
Here $F_A(q^2_{\perp})$ and $F_B(q^2_{\perp})$ are form
factors of the colliding mesons A
and B ($t\simeq -q^2_{\perp}$),
and $isP(s,q^2_{\perp})$ stands for the primary pomeron propagator
coupled to mesons $A$ and $B$. For diffractive processes the
main contribution is provided by the
imaginary part of the pomeron propagator;
 the real part of it may be neglected, although
in the calculation of scattering amplitudes the real part can be
easily restored.

For the description of data with
Lipatov's pomeron as a guide, the pomeron
propagator is parametrized  by a sum of several terms,
but in this Appendix, for the simplicity sake,  $P(s,q^2_{\perp})$ is
treated as  a one-pole
term, with $\alpha(0)=1+\Delta $. Then
$$
P(s,q^2_{\perp})= g_A g_B
s^{\Delta}e^{-q^2_{\perp}(2G+\alpha'\ln s)}.
\eqno{(A.2)}
$$
Within  the standard normalization, one has for the scattering
amplitude  $Im\,A(s,0)=s\sigma^{tot}$, though for the calculation of
multiple scatterings another amplitude normalization is more suitable,
namely, $ f(s,q^2_{\perp})= (is)^{-1}A(s,t) $.

We unitarize the upper and down blocks of the diffractive amplitudes
separately, and for this purpose
let us cut the pomeron amplitude
into two pieces as is shown in  Fig. 11b:
$$
f^{( P)}(s,q^2_\perp)= f_{A}^{( P)} (s_A,q^2_\perp)
f_{B}^{( P)} (s_B,q^2_\perp) \; .
\eqno{(A.3)}
$$
The upper and down pieces of the whole amplitude
depend on  the energies squared $s_A$ and $s_B$,
which  obey
 the equality $sm^2_0=s_As_B$  (below
$m_0=1$ GeV is chosen). Consider the upper block in details; it
is equal to:
$$
f_{A}^{( P)}(s_A,q^2_\perp)\ =\ F_A(q^2_\perp)g_A
s_A^{\Delta}e^{-(G+\alpha'\ln s_A)
q^2_\perp}\ .
\eqno{(A.4)}
$$
The amplitude $f_A^{( P)}(s_A,q^2_\perp)$ represented  as an
integral in the impact parameter space reads:
$$
f_{A}^{( P)}(s_A,q^2_\perp)=\int d^2b_A\,e^{i\vec q_\perp\vec b_A}
 \int d^2r_\perp\
\varphi_A^2(\vec r_\perp)\,\rho_A(\vec b_A-\vec r_\perp,s_A).
\eqno{(A.5)}
$$
Here $\rho_A(\vec b_A,s_A)$ is the pomeron propagator
coupled to  meson $A$,
$$
\rho_A(\vec b_A,s_A)=\int
\frac{d^2q_\perp}{(2\pi)^2}\ e^{-i\vec q_\perp\vec b_A}
\,g_A s_A^{\Delta}\, e^{-(G+\alpha'\ln s_A)q^2_\perp}\, ,
\eqno{(A.6)}
$$
while
$\varphi_A^2(\vec r_\perp)$
is the Fourier tranform of the  pion form factor:
$$
\varphi_A^2(\vec r_\perp)
=\int\frac{d^2q_\perp}{(2\pi)^2}\ e^{i\vec q_\perp\vec
r_\perp}\ F_A(q^2_\perp)\; .
\eqno{(A.7)}
$$
With this definition, the density
$\varphi_A^2(\vec r_\perp)$
 in the impact parameter space
is  invariant in respect to the boost along the $z$-axis.

The down block
of Fig. 11b is treated similarly, so one has:
$$
f_{B}^{( P)}(s_B,q^2_\perp)=\int d^2b_B\, e^{-i\vec q_\perp\vec b_B}
 \int d^2r\; '_\perp \;
\varphi_B^2(\vec r\; '_\perp)\,\rho_B(\vec b_B-\vec r\; '_\perp,s_B).
\eqno{(A.8)}
$$
Here we take into account that $\vec q_\perp$ is the incoming momentum
for the lower block, while for the upper block it is the outcoming one.

The one-pomeron exchange amplitude of Fig. 11a, after replacing
$\vec r_\perp\to \vec r$ and $\vec r\; '_\perp\to \vec r\; '$, reads:
$$
 f^{(P)}(s,q^2_\perp)=\int d^2b\, e^{i\vec q_\perp\vec b}
\int d^2 r\,\varphi^2_A(\vec r)\,d\vec r\;'\,\varphi^2_B(\vec r\;')
\,\chi(\vec r,\vec r\;',b),
\eqno{(A.9)}
$$
where
$$
{\chi}(\vec r,\vec r\;',b)=\int d^2b_A \,d^2b_B
\,\delta(\vec b-\vec b_A+\vec b_B)
\,\rho_A(\vec b_A-\vec r_\perp,s_A) \,\rho_B(\vec b_B-\vec r\;
'_\perp,s_B).
\eqno{(A.10)}
$$
Here $\chi(\vec r,\vec r \; ',b)$ is the eikonal profile
function, which takes account of all the multi-pomeron exchanges in
a standard way (for example, see \cite{eikonal,eikonal2}).
The amplitude with  $n$-pomeron exchanges shown in
 Fig. 3 ($n\ge 1$) is equal to:
$$
 f^{(PP...P)}(s,q^2_\perp)=\int d^2b\, e^{i\vec q_\perp\vec b}
\int d\vec r\,\varphi^2_A(\vec r)\,d\vec r\;'\,\varphi^2_B(\vec r\;')\,
\frac{2}{n!}\left(-\frac12 \chi(\vec r,\vec r\;',b )\right)^n\ .
\eqno{(A.11)}
$$
So, the scattering amplitude $AB\to AB$
 with a full set of the primary pomeron exchanges  reads:
$$
 f_{AB\to AB}(s,q^2_\perp)=2\int d^2b\, e^{i\vec q_\perp\vec b}
\int d\vec r\,\varphi^2_A(\vec r)\,d\vec r\;'\,\varphi^2_B(\vec r\;')
\left(1-e^{-\frac12 \chi (\vec r,\vec r\;',b)}\right) \,.
\eqno{(A.12)}
$$
The  normalization condition is $ f_{AB\to AB}(s,0)=\sigma^{tot}_{AB}$.

The equation (A.12) specifies neither the type of constituents
 responsible for  the pomeron interaction nor the  characteristics
of the constituent distributions,
$\varphi^2_A(\vec r)$ and $\varphi^2_B(\vec r)$,
 measured by the pomeron.
This specification will be done below in terms of the quark model.

{\bf b. Colliding meson as loosely bound $q\bar q$ system.}
Here the interpretation of the distribution
functions $\varphi^2_A(\vec r)$ and $\varphi^2_B(\vec r)$ is given in
terms of the quark model. For this purpose,  consider
the scattering process in the laboratory frame, initial
meson $A$ being at rest.  The form factor of the meson $A$ is
determined by the standard non-relativistic quark model expression:
$$ F_A(q_\perp)=\int \frac{d^3k}{(2\pi)^3} \psi_A(k) \psi_A(|\vec k+
\frac12\vec q_\perp|)=
\int d^3r_{q\bar q}
\phi^2_A(r_{q\bar q})e^{\frac i2 \vec r_{q\bar q}\vec q_\perp} \; ,
\eqno{(A.13)}
$$
 where $\vec k $ is
relative  quark--antiquark momentum,
 $ \vec k=\frac12( \vec k_q-\vec k_{ \bar q }) $,
and $\vec r_{q\bar q} $ is interquark
distance, $\vec r_{q\bar q}=\vec r_q-\vec r_{\bar q}$.  The
integration over $dr_{q\bar q z}$ introduces the quark density in the
$\vec r_{q\bar q \perp}$-space:
$$
\varphi_A^2(\vec r_\perp) = \int
dr_{q\bar q\, z}\phi^2_A(r_{q\bar q}).
\eqno{(A.14)}
$$
One more specification is
suitable here, namely,  an explicit integration over quark and
antiquark coordinates $\vec r_{q\;\perp}$ and $\vec r_{\bar q\;\perp}$:
$$
F_A(q_\perp)= \int
d^2\vec r_{q\,\perp} d^2\vec r_{\bar q\,\perp} \delta(\vec
r_{q\,\perp}+\vec r_{\bar q\,\perp})\Phi^2_A(\vec r_{q\,\perp}, \vec
r_{\bar q\,\perp}) e^{i \vec r_{q\,\perp} \vec q_{\perp}},
\eqno{(A.15)}
$$
where
$\Phi^2_A(\vec r_{q\perp},\vec r_{\bar q\perp})=
\varphi^2_A(\frac12|\vec r_{q\perp}-\vec r_{\bar q\perp}|)$.
Then, for the case  of a pomeron coupled to one quark of
the meson $A$ (Fig. 2a--b), the amplitude reads:
$$
f_{Aq}^{( P)} (s_A,q^2_\perp)=
\int d^2b_A e^{i\vec q_\perp\vec b_A}
\int d^2\vec r_{q\,\perp}
d^2\vec r_{\bar q\,\perp} \delta(\vec r_{q\,\perp}+\vec
r_{\bar q\,\perp})\Phi^2_A(\vec r_{q\,\perp},
\vec r_{\bar q\,\perp}) \rho_{Aq}(\vec
b_A-r_{q\,\perp},s_{Aq}) ,
\eqno{(A.16)}
$$
$$
\rho_{Aq}(\vec b_A,s_{Aq})=\int
\frac{d^2q_\perp}{(2\pi)^2}\ e^{-i\vec q_\perp\vec b_A}
\,g_{Aq}\, s_{Aq} ^{\Delta}\, e^{-(G+\alpha'\ln s_{Aq} )q^2_\perp}\, ,
$$
In  (A.16) we take into consideration that the quarks of the
mesons $A$ and $B$ share the invariant energy squared:
$$
    s_A =s_{Aq}+ s_{A\bar q}\; \;, \;\; s_B =s_{Bq}+ s_{B\bar q} \; .
\eqno{(A.17)}
$$
For mesons with equal quark masses
 $m_q =m_{\bar q}$,
$$
 s_{Aq}\simeq s_{A\bar q}\simeq \frac 12 s_A\,,\;\;\;{\rm and}\;\;\;
s_{Bq}\simeq s_{B\bar q}\simeq \frac 12 s_B \, .
\eqno{(A.18)}
$$
The interaction of meson $B$ with the primary
pomeron is treated in the same way,
implying the quark density in the impact parameter
space be invariant under the boost along  the $z$-axis:
$$
f_{Bq}^{( P)} (s_B,q^2_\perp)=
\int d^2b_B e^{-i\vec q_\perp\vec b_B}
\int d^2\vec r_{q\,\perp}
d^2\vec r_{\bar q\,\perp} \delta(\vec r_{q\,\perp}+\vec
r_{\bar q\,\perp})\Phi^2_B(\vec r_{q\,\perp},
\vec r_{\bar q\,\perp}) \rho_{Bq}(\vec
b_B-r_{q\,\perp},s_{Bq}) ,
\eqno{(A.19)}
$$
$$
\rho_{Bq}(\vec b_B,s_{Bq})=\int
\frac{d^2q_\perp}{(2\pi)^2}\ e^{i\vec q_\perp\vec b_B}
\,g_{Bq} \,s_{Bq} ^{\Delta}\, e^{-(G+\alpha'\ln s_{Bq} )q^2_\perp}\, ,
$$
 To take account of the
 pomeron exchanges between different quarks
(Fig. 2c), one should make a
substitution in  (A.16) and (A.19) as follows:
$$
\rho_{Aq}(\vec b_A-\vec r_{q\,\perp},s_{Aq})
\to
\rho_{Aq}(\vec b_A-\vec r_{q\,\perp},s_{Aq})
+\rho_{A\bar q}(\vec b_A-\vec r_{\bar q\,\perp},s_{A\bar q})
\equiv \rho_A^{(without\; cs)}(\vec b_A,\vec r ,s_A),
$$
$$
\rho_{Bq}(\vec b_B-\vec r\;_{q\,\perp}',s_{Bq})
\to
\rho_{Bq}(\vec b_B-\vec r\;_{q\,\perp}',s_{Bq})
+\rho_{B\bar q}(\vec b_B-\vec r\;_{\bar q\,\perp}',s_{B\bar q})
\equiv \rho_B^{(without\; cs)} (\vec b_B,\vec r\; ' ,s_B).
\eqno{(A.20)}
$$
Using (A.20), one can apply the formulae (A.10)--(A.12) to
the calculation of the scattering amplitude for $q\bar q$
mesons $A$ and $B$, with  compact notations for quark
variables:
$$
 d^2\vec r_{q\,\perp} d^2\vec r_{\bar q\,\perp}
\delta(\vec r_{q\,\perp}+
\vec r_{\bar q\,\perp})\Phi^2_A(\vec r_{q\,\perp},
\vec r_{\bar q\,\perp}) \equiv d^2 r \,\varphi^2_A(\vec r),
$$
$$
 d^2\vec r\;_{q\,\perp}' d^2\vec r\;_{\bar q\,\perp}'
\delta(\vec r\;_{q\,\perp}'+
\vec r\;_{\bar q\,\perp}')\Phi^2_B(\vec
r\;_{q\,\perp}', \vec r\;_{\bar q\,\perp}') \equiv d^2 r\; '
\,\varphi^2_B(\vec r \; ').
\eqno{(A.21)}
$$
This procedure provides us with a unitarized scattering amplitude,
but so far the colour screening has not been taken into consideration.

{\bf c. Colour screening effects for the scattering of loosely bound
systems.}
In the Lipatov's pomeron picture, the colour screening
is due to the two
types of coupling of  reggeized gluons to meson quarks,
either with the same quark (antiquark)
(Fig. 12a,b) or with both of them (Fig.  12c).

Let the meson $A$ be
at rest and $s_A$ rather large, $s_A \sim s$, see Fig. 12c; then
compare these two types of meson--pomeron vertices.
The amplitude related to the sum of the  Fig. 12a-b diagrams is
considered in more details in Appendix B; here we would like to
illustrate the scheme of how the colour screening emerges for a
loosely bound meson. The amplitude corresponding to diagrams
of Fig. 12a-c is equal to
$$
f^{({\bf P+P+PGG})}_A(q^2_\perp)=
2\int \frac{d^3 \kappa}{(2\pi)^3} \int \frac{d^3 k}{(2\pi)^3}
\left[ \psi_A(k)\psi_A(|\vec k+\frac12\vec q_\perp|) -
\psi_A(k)\psi_A(|\vec k+\vec \kappa_\perp|)\right]
$$
$$                                                 \times
a_{{\bf PGG}}\left( (\vec \kappa_{\perp} +\frac 12 \vec q_{\perp})^2,
(-\vec \kappa_{\perp} +\frac 12 \vec q_{\perp})^2, \kappa_z \right)\,.
\eqno{(A.22)}
$$
The term proportional to
$\psi_A(k)\psi_A(|\vec k+\frac12\vec q_\perp|) $ is the impulse
approximation contribution, while
the second one  in the integrand (A.22) is due to the triple-reggeon
diagram $a_{ {\bf PGG}}$, which itself is shown in Fig. 12c. It
depends on $s_A$ and
$\kappa_z =m_0M^2/s_A$, where $M^2$ is invariant energy squared
carried by the pomeron (see Appendix B for details):
$$
a_{{\bf PGG}}\left( (\vec \kappa_{\perp} +\frac 12 \vec q_{\perp})^2,
(-\vec \kappa_{\perp} +\frac 12 \vec q_{\perp})^2, \kappa_z \right)
\simeq R \; s_A^\Delta \; \kappa_z^{\alpha (0)-2\alpha_G(0)}\; .
\eqno{(A.23)}
$$
Here $\alpha_{{\bf P}}(0)$ and $\alpha_{{\bf G}}(0)$ are the pomeron and
reggeized gluon intercepts:
 $\alpha(0)=1+\Delta$ and $\alpha_{{\bf G}}(0) \simeq 1 $.
The coefficient
$R$ in the r.h.s.  (A.23) is a function of
$q^2_\perp $,  $(\vec \kappa_{\perp} +\frac 12 \vec q_{\perp})^2$ and
$(-\vec \kappa_{\perp} +\frac 12 \vec q_{\perp})^2$. Taking this
coefficient in the exponential form, as is usual for the reggeon
exchange amplitudes, one obtains:
$$
R\sim  e^{-\beta q^2_\perp }
e^{-\gamma (\vec \kappa_{\perp} +\frac 12 \vec q_{\perp})^2}
e^{-\gamma (-\vec \kappa_{\perp} +\frac 12 \vec q_{\perp})^2}=
e^{-(\beta+\frac12 \gamma) q^2_\perp}\;  e^{-2\gamma \kappa_{\perp}^2}.
\eqno{(A.24)}
$$

After having integrated over
 $ \vec k=\frac12( \vec k_q-\vec k_{ \bar q }) $, one has:
$$
f^{(P+P+PGG)}_A(q^2_\perp)=
2F_A(q^2_\perp )
\int \frac{d^3 \kappa}{(2\pi)^3}
a_{{\bf PGG}}\left( (\vec \kappa_{\perp} +\frac 12 \vec q_{\perp})^2,
(-\vec \kappa_{\perp} +\frac 12 \vec q_{\perp})^2, \kappa_z \right)$$
$$
-2\int \frac{d^3 \kappa}{(2\pi)^3} F_A(\kappa^2)
a_{{\bf PGG}}\left( (\vec \kappa_{\perp} +\frac 12 \vec q_{\perp})^2,
(-\vec \kappa_{\perp} +\frac 12 \vec q_{\perp})^2, \kappa_z \right)\,.
\eqno{(A.25)}
$$
This expression can be compared with (A.4) written
 for the impulse approximation amplitude; the comparison provides
$$
g_A s_A^{\Delta}e^{-(G+\alpha'\ln s_A)q^2_\perp}\ =
2\int \frac{d^3 \kappa}{(2\pi)^3}
a_{{\bf PGG}}\left( (\vec \kappa_{\perp} +\frac 12 \vec q_{\perp})^2,
(-\vec \kappa_{\perp} +\frac 12 \vec q_{\perp})^2, \kappa_z \right)\,.
\eqno{(A.26)}
$$
Equation (A.26) allows us to see the colour
screening in its explicit form:  for a point-like meson $A$ one has
$F_A=1$, and the amplitude (A.25) equals to zero. Still, more suitable
for this purpose is the
coordinate representation. The three-reggeon
amplitude $a_{{\bf PGG}}$ depends
on $\vec r_{g1}$ and $\vec r_{g2}$ which are the gluon coordinates
 in the impact parameter space:
$$
a_{{\bf PGG}}\left(
\kappa_{1\perp}^2, \kappa_{2\perp}^2, \kappa_z \right) = \int
d^2r_{g1}d^2r_{g2}\; a^{(coordinate)}_{{\bf PGG}}(r_{g1},r_{g2},\kappa_z)
\;e^{i(\vec r_{g1}\vec \kappa_{1\perp} +\vec r_{g2}\vec
\kappa_{1\perp})} \; .
\eqno{(A.27)}
$$
Then   (A.25) reads:
$$
f^{({\bf P+P+PGG})}_A(q^2_\perp)= 2\int \frac{d\kappa_z}{2\pi} \int
d^3r_{q\bar q} \int d^2r_{g1} d^2r_{g2}\; \phi^2_A (r_{q\bar q}) \;
a^{(coordinate)}_{{\bf PGG}}(r_{g1},r_{g2},\kappa_z)
$$
$$
\times\left[ e^{\frac i2
(\vec r_{q\bar q\perp} +\vec r_{g1} +\vec r_{g2})\vec q_\perp}
\delta (\vec r_{g1} -\vec r_{g2})
- e^{i r_{q\bar q z}\kappa_z}
e^{\frac i2(\vec r_{g1} +\vec r_{g2})\vec q_\perp}
\delta (\vec r_{g1} -\vec r_{g2} - \vec r_{q\bar q \perp}) \right ]
\; .
\eqno{(A.28)}
$$
The integrand in  (A.28) tends to zero with
$| \vec r_{q\bar q \perp}| \to 0$ and $r_{q\bar q z} \to 0$: this  is
a manifestation of the colour screening. Moreover, the colour screening
reveals itself when, after the integration over $ r_{q\bar q z}$,
the expression    $| \vec r_{q\bar q \perp}|$ tends to zero.
 The matter is that the
dominant contribution to the integral (A.28) is given by the region
$ \kappa_z \sim 0$. Hence
$$
| r_{q\bar q z}\kappa_z| \ll 1\,.
\eqno{(A.29)}
$$
 So, with a sufficiently good accuracy, one can substitute
 in  (A.28)\\ $\exp(i r_{q\bar q
z}\kappa_z) \to 1$ (for more details see Appendix  B).
 As a result, using the variables
$\vec b_A =\frac 12 (\vec r_{g1}+\vec r_{g2})$ and $\vec r_{gg}=
\vec r_{g1} -\vec r_{g2} $, we have
$$f^{({\bf P+P+PGG})}_A(q^2_\perp)=$$
$$2\int \frac{d\kappa_z}{2\pi} \int d^2r_{q\bar q \perp}
\int d^2b_A
\varphi^2_A (r_{q\bar q \perp}) \;
a^{(coordinate)}_{{\bf PGG}}(b_A^2,b_A^2,\kappa_z)e^
{i (\vec b_A + \frac 12 \vec r_{q\bar q \perp}) \vec q_{\perp} }
$$
$$
-2\int \frac{d\kappa_z}{2\pi} \int d^2r_{q\bar q \perp}
\int d^2b_A \varphi^2_A (r_{q\bar q \perp}) \;
a^{(coordinate)}_{{\bf PGG}}((\vec b_A -\frac 12 \vec r_{q\bar q \perp})^2 ,
(\vec b_A +\frac 12 \vec r_{q\bar q \perp})^2 , \kappa_z)
e^ {i \vec b_A \vec q_{\perp}}.
\eqno{(A.30)}
$$
Finally, with the re-definition in the first term
$\vec b_A + \frac 12 \vec r_{q\bar q \perp} \to \vec b_A $, we have:
$$f^{({\bf P+P+PGG})}_A(q^2_\perp)=$$
$$
2 \int d^2b
e^ {i \vec b_A \vec q_{\perp}}
 \int d^2r_{q\bar q \perp}
\varphi^2_A (r_{q\bar q \perp}) \;
\int \frac{d\kappa_z}{2\pi}
a^{(coordinate)}_{{\bf PGG}}
\left ( (\vec b_A - \frac 12 \vec r_{q\bar q \perp})^2,
(\vec b_A - \frac 12 \vec r_{q\bar q \perp})^2, \kappa_z \right )
$$
$$
-2 \int d^2b_A
e^ {i \vec b_A \vec q_{\perp}}.
\int d^2r_{q\bar q \perp}
 \varphi^2_A (r_{q\bar q \perp}) \;
\int \frac{d\kappa_z}{2\pi}
a^{(coordinate)}_{{\bf PGG}}((\vec b_A -\frac 12 \vec r_{q\bar q \perp})^2 ,
(\vec b_A +\frac 12 \vec r_{q\bar q \perp})^2 , \kappa_z)\,.
\eqno{(A.31)}
$$
In our notations
$$
\rho_A(\vec b_A, s_A) =
\int \frac{d\kappa_z}{2\pi}
a^{(coordinate)}_{{\bf PGG}}( b_A^2, b_A^2, \kappa_z),
\eqno{(A.32)}
$$
then with the exponential parametrization (A.23) one has:
$$
\int \frac{d\kappa_z}{2\pi}
a^{(coordinate)}_{{\bf PGG}}
\left ( (\vec b_A -\frac 12 \vec r_{q\bar q \perp})^2 ,
(\vec b_A +\frac 12 \vec r_{q\bar q \perp})^2 , \kappa_z \right )
=\rho_A (\vec b, s_A)
 \exp\left ( -\frac {r_{q\bar q \perp}^2}{r^2_{cs}} \right ) \,.
\eqno{(A.33)}
$$
Using the variables given in  (A.15)--(A.16), we have
$$
f^{(P+P+PGG)}_A(q^2_\perp)= 2 \int d^2b
e^ {i \vec b_A \vec q_{\perp}}
\int d^2\vec
r_{q\,\perp} d^2\vec r_{\bar q\,\perp} \delta(\vec r_{q\,\perp}+\vec
r_{\bar q\perp})\Phi^2_A(\vec r_{q\perp},
\vec r_{\bar q\perp})
\rho_A (\vec b_A, \vec r , s_A)\,,
\eqno{(A.34)}
$$
with the colour sceening term
 included into the primary pomeron amplitude:
$$\rho_A (\vec b_A, \vec r , s_A) =
 \rho_A  ( \vec b_A - \vec r_{q\perp}, s_A) +
\rho_A ( \vec b_A- \vec r_{\bar q\perp}, s_A)
 -2 \rho_A (\vec b -\frac {\vec r_{q \perp}+\vec r_{\bar q \perp}}{2} ,
s_A)
e^{ -\frac{(\vec r_{q \perp}-\vec r_{\bar q \perp})^2}{4r^2_{cs}} }.
\eqno{(A.35)}
$$
This amplitude should be compared with (A.20), written without
colour screening term.

 Likewise,
the amplitude $f^{({\bf P+P+PGG})}_B(q^2_\perp)$ is written,
 with the  replacements $A \to B $ and
$\vec q_\perp \to -\vec q_\perp $.

The full amplitude with the $s$-channel unitarization
is given by  (A.12) with the profile function determined as follows:
$$
{\chi}(\vec r,\vec r\;',b)=\int d^2b_A\, d^2b_B\,
\delta(\vec b-\vec b_A+\vec b_B)
\,\rho_A(\vec b_A, \vec r, s_A)\, \rho_B(\vec b_B, \vec r\;' ,s_B).
\eqno{(A.36)}
$$

\section*{Appendix B. Meson--pomeron coupling}

Here we calculate the coupling of the
three-reggeon amplitude {\bf PGG} to meson. The meson is treated
as a loosely bound $q\bar q$ system.

{\bf a. Three-reggeon amplitude $a_{{\bf PGG}}$}. This amplitude depends on
three invariant energies squared which are
rather large, $s$, $s'$ and
$M^2$, and three momentum transfers, $t_1$, $t_2$ and
$q^2$, which are small (see Fig. 12c). With standard normalization, the
amplitude {\bf PGG} has the form \cite{eikonal,A2}:
$$
A_{{\bf PGG}}=R(t_1,t_2,q^2)e^{i\frac{\pi}{2}\alpha_{{\bf P}} (q^2)}
(M^2)^{\alpha_{{\bf P}}(q^2)}
\left ( \frac {s}{M^2}\right )^{\alpha_{{\bf G}}(t_1)}
\left (\frac {s'}{M^2}\right )^{\alpha_{{\bf G}}(t_2)}.
\eqno{(B.1)}
$$
 At $\alpha_{{\bf P}} (0) \simeq 1 $, the imaginary part of the
 pomeron amplitude provides a domiminant contribution.
 Then, using an exponential parametrisation for the momentum
transfer dependence, one has:
$$
A_{{\bf PGG}} \simeq
i\,R\,e^{-\beta(\kappa^2_{1\perp}+\kappa^2_{2\perp})-\gamma
q^2_{\perp}}\,(ss')^{\frac{\alpha_{{\bf P}}(0)}{2}}
(yy')^{\frac{\alpha_{{\bf P}}(0)}{2}- \alpha_{{\bf G}}(0)}.
\eqno{(B.2)}
$$
Here $t_i
\simeq -\kappa^2_{i\perp}$ and $q^2 \simeq -q^2_{\perp}$. The
following notations are used: $y=M^2/s$ and $y'=M^2/s'$.
For the three-reggeon amplitude, $y$ and $y'$ are
small, because $s' \sim   s << M^2$. The coefficients $\beta$ and
$\gamma$ include the weak (logarithmic) dependence on $s$, $s'$ and
$M^2$ that originates from the
standard expansion of reggeon trajectories:
   $\alpha_{{\bf P}}(q^2) \simeq \alpha_{{\bf P}}(0)-
\alpha'_{{\bf P}}q^2_\perp$ and $\alpha_{{\bf G}}(t) \simeq
\alpha_{{\bf G}}(0)- \alpha'_{{\bf G}}\kappa^2_\perp$.

 Imposing $s'=s$, the amplitude used in calculations is
 as follows:
$$
a_{{\bf PGG}}=\frac1{is}A_{{\bf PGG}} \simeq
R\,e^{-\beta(\kappa^2_{1\perp}+\kappa^2_{2\perp})-\gamma q^2_{\perp}}\\
s^{\Delta} y^{\alpha_{{\bf P}}(0)-2\alpha_{{\bf G}}(0)}\,.
\eqno{(B.3)}
$$
Being a real function,
the amplitude  $a_{{\bf PGG}}$
is related to cutting of Fig. 12d--diagram
along the pomeron line. This means that
$sa_{{\bf PGG}}$ given by (B.3) is a discontinuity of $A_{{\bf PGG}}$ across
the $M^2$-cut:
$$a_{{\bf PGG}}\simeq \frac1{s}disc_{M^2}A_{{\bf PGG}}.
\eqno{(B.4)}
$$
Therefore,
 $A_{{\bf PGG}}$  can be represented as a dispersion integral over $M^2$,
 $sa_{{\bf PGG}}$ being an integrand.

{\bf b. Two-gluon interaction with a
quark}. The interaction diagram is shown in Fig. 12a: reggeized
gluons interact with the same quark. The meson A is treated in the
rest frame, therefore, a non-relativistic
quark propagator technique  is appropriate here:
$(m^2-k^2)^{-1} \simeq (-2mE+\vec k^2-i0)^{-1}$, where
$E=k_0-m$. The amplitude of the Fig. 12a--diagram is:
$$
 A^{(P)}(s,q^2_{\perp})=\int \frac{dE_{\bar q}
d^3k_{\bar q}}{i(2\pi)^4}\, \int  \frac{d^4\kappa_1}{i(2\pi)^4}
  \frac{G_A}{-2mE_{ q}+\vec
k^2_{ q}-i0}\cdot \frac1{-2mE_{\bar q}+ \vec k_{\bar q}^2-i0}
$$
$$
\times \frac1{-2mE''_q+ \vec
k''^2_q-i0}
\cdot \frac{G_A}{-2mE'_q+\vec k'^2_q-i0}  \int\ \frac{dM^2}{\pi} \cdot
\frac {g^2 \tilde a
(\kappa^2_1,\kappa^2_2,M^2)}{M^2-(P_B-\kappa_1)^2-i0} .
\eqno{(B.5)}
$$
The notations of momenta are shown in Fig. 12a. The
vertex function
$G_A$ depends on the relative quark--antiquark momentum, namely,
$(k_q-k_{\bar q})^2 \simeq  -(\vec k_q-\vec k_{\bar q})^2$ for the
incoming meson vertex and $(\vec k'_q-\vec k_{\bar q})^2 \simeq
-(\vec k'_q-\vec k_{\bar q})^2$ for outgoing one.
The three-reggeon amplitude ${\bf PGG}$ is written as a dispersion
integral over $M^2$, $g$ being the
quark-gluon coupling. Three integrations in  (B.1) are easy to perform.
When integrating
over  $E_{\bar q}$ and $\kappa_0$, the substitutions are made:
$$
(-2mE_{\bar q}+\vec k^2_{\bar q}-i0)^{-1} \to \frac {i\pi}{m}
\delta\left (E_{\bar q}-\frac {\vec k^2_{\bar q}}{2m}
\right ),
$$
$$ (-2mE''_q+\vec
k''^2_q-i0)^{-1}
 \to \frac {i\pi}{m} \delta\left (\kappa_{10}-\epsilon-\frac
{(\vec k^2_q+\vec \kappa_1)^2}{2m}
-\frac {\vec k^2_{\bar q}}{2m^2}\right ).
\eqno{(B.6)}
$$
where
$\epsilon =2m-\mu_A$. The real part of the three-regeon diagram
$A_{{\bf PGG}}$ is small, see  (B.4). Therefore, in the dispersion
integral over $M^2$ the main contribution comes from the half-residue:
  $$
\left
(M^2-(P_B-\kappa_1)^2 -i0 \right )^{-1} \to i\pi  \delta\left
(M^2 +2p_B\kappa_{1z}\right )\,,
\eqno{(B.7)}
$$
$p_B$ being a large momentum carried by a particle $B$ along the
$z$-axis:  $P_B= (P_{B0},\vec P_{\perp},P_{Bz})= (p_B+
\mu_B^2/(2p_B),0,-p_B)$ and
$s \simeq
2p_B \mu_A$. The terms of the order of $m/p_B$ are neglected.

After integrating over $E_{\bar q}$, $\kappa_{10}$ and
$M^2$, we obtain:
$$
A^{(P)}(s,q^2_\perp)=\int \frac{d^3k_{q\bar
q}}{(2\pi)^3}\psi_A(k_{q\bar q})\psi_A(|\vec k_{q\bar q}+
\frac12\vec q_{\perp}|)\, \int
\frac{d^3\kappa_1}{(2\pi)^3}\,
\frac{i}{2m}\,
g^2\, \tilde a (\kappa^2_{1\perp},\kappa^2_{2\perp},-2p_B\kappa_{1z}),
\eqno{(B.8)}
$$
where
$$
\psi_A(k_{q\bar q})=\frac{G_A}
{2\sqrt{2m}(m\epsilon +\vec k^2_{q\bar q})}.
\eqno{(B.9)}
$$
For the three-reggeon diagram of Fig. 12a the
following constraint is imposed:
$s>>M^2=-2p_B\kappa_{1z}>>m^2$. This means that $\kappa_{1z}$ is
negative and small, $|\kappa_{1z}|\ll \mu_A$.

Comparing  (B.8) with the first term of the right-hand
side of (A.22) gives us the following equality:
$$
g^2\, \tilde a (\kappa^2_{1\perp},\kappa^2_{2\perp},2p_B\kappa_z)
=2m\,s\, a_{{\bf PGG}}
(\kappa^2_{1\perp},\kappa^2_{2\perp},\kappa_z)F_B(q^2_\perp) \,,
\eqno{(B.10)}
$$
where $\kappa_{1z}=-\kappa_z$.
Here we take into account that the form factor  of meson B,
$F_B(q^2_\perp)$, enters the lower block of Fig. 12a.

The expression for the antiquark--pomeron interaction is identical to
that of the  quark, for the
integrand (B.8) is invariant with respect to the  replacement
$\vec k_{q} \to \vec k_{\bar q}$ and $g \to -g$.

{\bf c. The interaction of gluons with quark and antiquark.}
 The graphical representation of the amplitude is
  shown in Fig.  12b. The amplitude reads:
$$
 A^{({\bf PGG})}(s,q^2_{\perp})=\int \frac{dE_{\bar q}
d^3k_{\bar q}}{i(2\pi)^4}\, \int  \frac{d^4\kappa_1}{i(2\pi)^4}
  \frac{G_A}{-2mE_{\bar q}+\vec
k^2_{\bar q}-i0} \cdot \frac1{-2mE_q+\vec k^2_q-i0}$$
$$
\times \frac1{-2mE'_{\bar q}+
\vec k'^2_{\bar q}-i0}
\cdot\frac{G_A}{-2mE'_q+ \vec k'^2_q-i0}
\int\ \frac{dM^2}{\pi} \cdot
\frac {(-g^2) \; \tilde a
(\kappa^2_{1\perp},\kappa^2_{\bar q},M^2)}{M^2-(P_B-\kappa_1)^2-i0} .
\eqno{(B.11)}
$$
The factor $-g^2$ in the r.h.s. of (B.11)
is due to  the interaction of gluons with  quark and antiquark.
The integrations over $E_{\bar q}$ and $E'_q$
are equivalent to the replacements:
$$
(-2mE_{\bar q}+\vec k^2_{\bar q})^{-1} \to
\frac {i\pi}{m}
\delta \left (E_{\bar q}-\frac {\vec k^2_{\bar q}}{2m} \right ),
$$
$$ (-2mE'_q+\vec k'^2_q)^{-1}  \to
\frac {i\pi}{m} \delta \left ( \epsilon+ \frac {\vec k^2_{\bar q}}{2m^2}
+\frac {(\vec k^2_q+\vec \kappa_1)^2}{2m} -\kappa_{10} \right ),
\eqno{(B.12)}
$$
and the integration over $M^2$ is eliminated due to the
replacement (B.7). We have
$$
A^{({\bf PGG})}(s,q^2_\perp)=-\int \frac{d^3k_{\bar q}}{(2\pi)^3}\,
\int \frac{d^3\kappa}{(2\pi)^3}
\frac {G_A}{2\sqrt{2m} (m\epsilon +\vec k^2_{q\bar q}) } \cdot
\frac {G_A}{2\sqrt{2m} (m\epsilon +(\vec k_{q\bar q}+\vec \kappa)^2) }
$$
$$
\times \frac {i}{m}\, g^2\,
\tilde a(\kappa^2_{1\perp},\kappa^2_{2\perp},2p_B\kappa_z)\,.
\eqno{(B.13)}
$$
Implying  (B.9) and (B.10), one has the second term in (A.22).

Here, as for the diagram of Fig. 12a, the
three-regeon amplitude determines the integration region over
$\kappa_z$:  the constraint $s\gg M^2$ means that $\kappa_z $ is small
in the hadronic scale.  This justifies the approximation given by
(A.29).

\section*{Appendix C. Pomeron--meson interaction in terms of the
light--cone variables}

Here, in terms of the light--cone variables, we calculate  the diagrams
shown in Fig. 12a,b.

{\bf a. The pomeron interacting with a quark, Fig. 12a.}
This diagram  written as a spectral
integral over $q\bar q$ invariant mass reads:
  $$
A^{(P)}(s,q^2_\perp)=\int_{4m^2}^\infty
\frac{dM^2_{q\bar q} dM''^2_{q\bar q}dM'^2_{q\bar q}}{\pi^3}
\cdot\frac{dM^2}{\pi}\frac{d^4\kappa_1}{i(2\pi)^4}
\cdot
\frac{G_A (M^2_{q\bar q})
d\Phi_2(P;k_q,k_{\bar q})}{M^2_{q\bar q}-\mu^2_A}
$$
$$
\times
 \frac{d\Phi_1(P'';k''_q,k_{\bar q})}
{M''^2_{q\bar q}-(P_A+\kappa_1)^2-i0} \cdot
  \frac{d\Phi_1(P';k'_q,k_{\bar q})
G_A(M'^2_{q\bar q}) }{M'^2_{q\bar q}-\mu^2_A}
S^{(q)}_I
\frac{g^2 \hat a(\kappa^2_{1\perp},\kappa^2_{2\perp},M^2)}
{M^2-(P_B-\kappa^2_1)^2-i0}\,.
\eqno{(C.1)}
$$
The detailed presentation of the spectral integration technique
and its application to the
description of  composite systems can be found
 in \cite{AMN}.
$G_A$ is the
vertex function for the transition $meson \; A \to q\bar q$;
$d\Phi_2$ and $d\Phi_1$
are the phase spaces of the $q\bar q$ intermediate states:
$$
d\Phi_2(P;k_q,k_{\bar q})=\frac12  \frac{d^3k_q}{(2\pi)^32k_{q0}}
\frac{d^3k_{\bar q}}{(2\pi)^32k_{\bar q}} (2\pi)^4
\delta^4(P-k_q-k_{\bar q}),$$
$$
d\Phi_1(P';k'_q,k_{\bar q})=\frac12  \frac{d^3k'_q}{(2\pi)^32k'_{q0}}
\delta^4(P'-k'_q-k_{\bar q})\,.
\eqno{(C.2)}
$$
Here $P$ and $P'$ stand for the total four-momenta of the intermediate
states
with invariant masses $M_{q\bar q}$ and $M'_{q\bar q}$:
$P^2 =M^2_{q\bar q}$ and $P'^2 = M'^2_{q\bar q}$.
To introduce the light--cone
variables, the centre-of-mass system
of the colliding particles $A$ and $B$ is the most suitable.
Using  the momentum notation $p=(p_0, \vec
p_\perp, p_z)$, one has for $P_A$ and $P_B$:
$$
P_A=(p+\frac {\mu^2_A}{2p}, 0,p),\,\,\,
P_B=(p+\frac {\mu^2_B}{2p}, 0,-p).
\eqno{(C.3)}
$$
The factor  $S_I^{(q)} $ is defined by the spin variables
of quarks:
$$
S_I^{(q)} = -{\mbox Sp}\left [ \Gamma_A\,(\hat k'_1 +m)\,\hat n \,
(\hat k''_1 +m)\,\hat n \,
(\hat k_1 +m)\, \Gamma_A\, (-\hat k_2 +m) \right ]\,,
\eqno{(C.4)}
$$
where $\Gamma_A$ is the spin-dependent factor for the vertex $ A \to
q\bar q$, for example, $\Gamma_A=\gamma_5$ for meson,
 and $\hat n$ quark--gluon vertex:
$$
\hat n=\gamma_\alpha n_\alpha , \; \;
 n=\frac1 {2p}(1,0,-1).
\eqno{(C.5)}
$$
The gluon polarization $n_\alpha$ which is parallel to $P_B$
provides the main contribution into fermion loop related to meson A
\cite{GGF}.

The phase space factors in terms
of  light--cone variables read:
$$
d\Phi_2(P;k_q,k_{\bar q})=\frac1{(4\pi)^2} \frac {dx_q dx_{\bar q}}
{x_qx_{\bar q}}\delta(1-x_q-x_{\bar q})
d^2k_{q\perp}d^2k_{\bar q\perp}\delta(
\vec k_{q\perp}+\vec k_{\bar q\perp})\delta
\left (
M^2_{q\bar q}-\frac{m_{q\perp}^2}{x_q}-\frac{m_{\bar q\perp}^2}
{ x_{\bar q}}\right) \,,  $$
$$
d\Phi_1(P';k'_{ q},k_{\bar q})=
\pi \frac {dx'_q }{x'_q}\delta(1-x'_q-x_{\bar q})
d^2k'_{q\perp}
\delta(\vec k'_{q\perp}+\vec k_{\bar q\perp}-\vec q_\perp)
\delta\left (
M'^2_{q\bar q}+\vec q^{\;2}_\perp-\frac{m'^2_{q\perp}}{x'_q}-
\frac{m_{\bar q\perp}^2}{x_{\bar q}}
\right) \,,$$
$$
d\Phi_1(P'';k''_q,k_{\bar q})=
\pi \frac {dx''_q }{x''_q}\delta(1-x''_q-x_{\bar q})
d^2k''_{q\perp}
\delta(\vec k''_{q\perp}+\vec k_{\bar q\perp}-\vec \kappa_{1\perp})
\delta\left (
M''^2_{q\bar q}+\vec
 \kappa_{1\perp}^{\; 2}-\frac{m''^2_{q\perp}}{x''_q}-
\frac{m_{\bar q\perp}^2}{x_{\bar q}}
\right) \,.
\eqno{(C.6)}
$$
Here $x_q=k_{qz}/p$ and $m^2_{q\perp}=m^2+k^2_{q\perp}$.
An important point is that
$\kappa_z/p$ is small at large $p$: it follows from the constraint
$M^2 \ll s$ for three-reggeon diagrams. The integration
over $M^2_{q\bar q}$, $M'^2_{q\bar q}$, $M''^2_{q\bar q}$
  and  the  substitution
$\left (M^2-(P_N-\kappa_1)^2-i0\right )^{-1} \to
i\pi\delta\left (M^2 -(P_B-\kappa_1)^2 \right )$ (see  (C.7))
give:
$$
A^{(P)}(s,q^2_\perp)=\frac1 {4\pi}\int_0^1 \frac {dx}{x(1-x)^3}
\int \frac {d^2k_\perp}{(2\pi)^2}\cdot
\frac {G_A(M^2_{q\bar q})}
{M^2_{q\bar q}-\mu_A^2}\frac {G_A(M'^2_{q\bar q})}
{M'^2_{q\bar q}-\mu_A^2}
$$
$$\times \int \frac{d\kappa_{10}d\kappa_{1z}d^2\kappa_{1\perp}}
{(2\pi)^4}S_I^{(q)}
\frac{g^2
\hat a_{{\bf PGG}}\left (\kappa^2_{1\perp},\kappa^2_{2\perp},
-2p(\kappa_{10}+\kappa_{1z})\right ) }
{M''^2_{q\bar q}-2p(\kappa_{10}-\kappa_{1z})-i0}
  \; ,
\eqno{(C.7)}
$$
where $x_{\bar q} \equiv x$,
$\vec k_{\bar q \perp} \equiv k_{ \perp}$ and
$$
M^2_{q\bar q}
=\frac {m^2+k^2_\perp}{x(1-x)},\;\;\;
M'^2_{q\bar q}
=\frac {m^2+(\vec k_\perp-x\vec q_\perp)^2}{x(1-x)},\;\;\;
M''^2_{q\bar q}
=\frac {m^2+(\vec k_\perp-x\vec \kappa _{1\perp})^2}{x(1-x)}.
\eqno{(C.8)}
$$
The integration over $\kappa_-=\kappa_{10}-\kappa_{1z}$ is equivalent
to the substitution
$(M''^2_{q\bar q}
-2p\kappa_--i0)^{-1} \to
\frac{i\pi}{p}\delta(\kappa_-)$,
so we have
$$
A^{(P)}(s,q^2_\perp)=\frac1 {4\pi}\int_0^1 \frac {dx}{x(1-x)^3}
\int \frac {d^2k_\perp}{(2\pi)^2}\cdot
\frac {G_A(M^2_{q\bar q})}
{M^2_{q\bar q}-\mu_A^2}
\frac {G_A(M'^2_{q\bar q})}{M'^2_{q\bar q}-\mu_A^2}
\; S_I^{q\bar q}
 $$
$$\times
\int \frac{d\kappa_{+}d^2\kappa_{\perp}}
{(2\pi)^3}\cdot \frac{ig^2}{4p} \;
\hat a_{{\bf PGG}}\left (\kappa^2_{1\perp},\kappa^2_{2\perp},
2p\kappa_{+})\right ).
\eqno{(C.9)}
$$
The factor $G_A(M^2_{q\bar q})/(M^2_{q\bar q}-\mu_A^2)$
 determines the
wave function of meson $A$. The pion vertex $G_\pi(M^2_{q\bar
q})$ has been found in  \cite{AMN} from the experimental data on
 pion form factor.

The factor $S_I^{q}$ is given by  (C.4).
One can re-write it, using the
equality $\hat n \hat n=0$, as follows:
$$
S_I^{( q)} =-
2x_q{\mbox Sp}\left [ \Gamma_A(\hat k'_q +m)\hat n
(\hat k_q +m)\Gamma_A (-\hat k_{\bar q} +m) \right ]\,,
\eqno{(C.10)}
$$
 $\hat n$ being a spin-dependent quark--pomeron vertex.

{\bf b. The pomeron interacting with antiquark.}
When two reggeized gluons interact with antiquark,
the spin-dependent part of the loop diagram reads:
$$ S_{I}^{(\bar q)} = -{\mbox Sp}\left [ \Gamma_A\,(\hat k_q+m)\,
\Gamma_A\, (-\hat k_{\bar q} +m)\,\hat n\,
(-\hat k''_{\bar q} +m)\, \hat n \,(-\hat k'_{\bar q} +m)
\right ]$$
$$=-
2x_{\bar q}{\mbox Sp}\left [ \Gamma_A
(\hat k_q +m)\Gamma_A (-\hat k_{\bar q} +m)
(-\hat n)(-\hat k'_{\bar q} +m)\right ].
\eqno{(C.11)}
 $$
The  antiquark--pomeron vertex is equal to $-\hat n$.

The momentum-dependent part of the loop diagram is determined by
 (C.9), with the re-definitions $x \to (1-x)$ and $\vec k_\perp
\to \; -\vec k_\perp $.

{\bf c. The pomeron interacting with quark and antiquark, Fig. 12b.}
The diagram of Fig. 12b written as a spectral
integral over the $q\bar q$ invariant mass reads:
  $$
A^{(P)}(s,q^2_\perp)=\int_{4m^2}^\infty
\frac{dM^2_{q\bar q} dM''^2_{q\bar q}dM'^2_{q\bar q}}{\pi^3}\cdot
\frac{dM^2}{\pi}\frac{d^4\kappa_1}{i(2\pi)^4}
\cdot
\frac{G_A (M^2_{q\bar q})
d\Phi_2(P;k_q,k_{\bar q})}{M^2_{q\bar q}-\mu^2_A}
$$
$$\times
\frac{ d\Phi_1(P'';k'_q,k_{\bar q})}
{M''^2_{q\bar q}-(P_A+\kappa_1)^2-i0} \cdot
\frac{d\Phi_1(P';k'_q,k'_{\bar q})
G_A(M'^2_{q\bar q}) }{M'^2_{q\bar q}-\mu^2_A}
S_{II}
\frac{g^2 \hat a(\kappa^2_{1\perp},\kappa^2_{2\perp},M^2)}
{M^2-(P_B-\kappa^2_1)^2-i0}
\eqno{(C.12)}
$$
The factor  $S_{II} $ is defined by the quark spin variables:
$$
S_{II}^{(q)} = -{\mbox Sp}\left [ \Gamma_A\,(\hat k'_1 +m)\,\hat n\,
 (\hat k_1 +m)\, \Gamma_A\, (-\hat k_2 +m) \,\hat n\,
(-\hat k'_2 +m)\right ].
\eqno{(C.13)}
$$
The phase space factors in terms
of the  light--cone variables read:
$$
d\Phi_2(P;k_q,k_{\bar q})=\frac1{(4\pi)^2} \frac {dx_q dx_{\bar q}}
{x_qx_{\bar q}}\delta(1-x_q-x_{\bar q})
d^2k_{q\perp}d^2k_{\bar q\perp}\delta(
\vec k_{q\perp}+\vec k_{\bar q\perp})\delta
\left (
M^2_{q\bar q}-\frac{m_{q\perp}^2}{x_q}-\frac{m_{\bar q\perp}^2}{x_{
\bar q}}
\right) \,,  $$
$$
d\Phi_1(P';k'_{ q},k'_{\bar q})=
\pi \frac {dx'_{\bar q} }{x'_{\bar q}}\delta(1-x'_q-x'_{\bar q})
d^2k'_{\bar q \perp}
\delta(\vec k'_{q\perp}+\vec k'_{\bar q\perp}-\vec q_\perp)
\delta\left (
M'^2_{q\bar q}+\vec q^{\;2}_\perp-\frac{m'^2_{q\perp} }{x'_q}-
\frac{m'^{2}_{\bar q\perp}}{x'_{\bar q} }
\right) \,,
$$
$$
d\Phi_1(P'';k'_q,k_{\bar q})=
\pi \frac {dx'_q }{x'_q}\delta(1-x'_q-x_{\bar q})
d^2k'_{q\perp}
\delta(\vec k'_{q\perp}+\vec k_{\bar q\perp}-\vec \kappa_{1\perp})
\delta\left (
M''^2_{q\bar q}+\vec
 \kappa_{1\perp}^{\; 2}-\frac{m'^2_{q\perp}}{x'_q}-
\frac{m_{\bar q\perp}^2}{x_{\bar q}}
\right) \,.
\eqno{(C.14)}
$$
Integrating over $M^2_{q\bar q}$,
$M'^2_{q\bar q}$,
$M''^2_{q\bar q}$  and $M^2$,
one gets:
$$
A^{(P)}(s,q^2_\perp)=\frac1 {4\pi}\int_0^1 \frac {dx}{x^2(1-x)^2}
\int \frac {d^2k_\perp}{(2\pi)^2}\cdot
\frac {G_A(M^2_{q\bar q})}
{M^2_{q\bar q}-\mu_A^2}\frac {G_A(M'^2_{q\bar q})}
{M'^2_{q\bar q}-\mu_A^2}
$$
$$\times \int \frac{d\kappa_{10}d\kappa_{1z}d^2\kappa_{\perp}}
{(2\pi)^4}  S_{II}
\frac{(-g^2)
\hat a_{{\bf PGG}}\left (\kappa^2_{1\perp},\kappa^2_{2\perp},
-2p(\kappa_{10}+\kappa_{1z})\right ) }
{M''^2_{q\bar q}-2p(\kappa_{10}-\kappa_{1z})-i0}
  \; .
\eqno{(C.15)}
$$
The invariant masses squared in
the intermediate states are:
$$
M^2_{q\bar q}
=\frac {m^2+k^2_\perp}{x(1-x)},\;\;\;
M'^2_{q\bar q}
=\frac {m^2+(\vec k_\perp-x\vec q_\perp)^2}{x(1-x)},\;\;\;
M''^2_{q\bar q}
=\frac {m^2+(\vec k_\perp-x\vec \kappa _{1\perp}
-(1-x)\vec \kappa_{2\perp})^2}{x(1-x)}.
\eqno{(C.16)}
$$
The integration over $\kappa_-=\kappa_{10}-\kappa_{1z}$ is equivalent
to the substitution $(M''^2_{q\bar q} -2p\kappa_--i0)^{-1} \to
\frac{i\pi}{p}\delta(\kappa_-)$,
so we have
$$
A^{(P)}(s,q^2_\perp)=-\frac1 {4\pi}\int_0^1 \frac {dx}{x(1-x)^3}
\int \frac {d^2k_\perp}{(2\pi)^2}\cdot
\frac {G_A(M^2_{q\bar q})}
{M^2_{q\bar q}-\mu_A^2}
\frac {G_A(M'^2_{q\bar q})}{M'^2_{q\bar q}-\mu_A^2}
 $$
$$\times
\int \frac{d\kappa_{+}d^2 \kappa_{\perp}}
{(2\pi)^3}\cdot S_{II} \frac{ig^2}{4p} \;
\hat a_{{\bf PGG}}\left (\kappa^2_{1\perp},\kappa^2_{2\perp},
2p\kappa_{+})\right ).
\eqno{(C.17)}
$$
 Here, in line with  (C.10), the sign of $\kappa_{+}$ is changed:
$\kappa_{+}=-\kappa_{10}- \kappa_{1z}$.

\newpage
\begin{figure}
\centerline{\epsfig{file=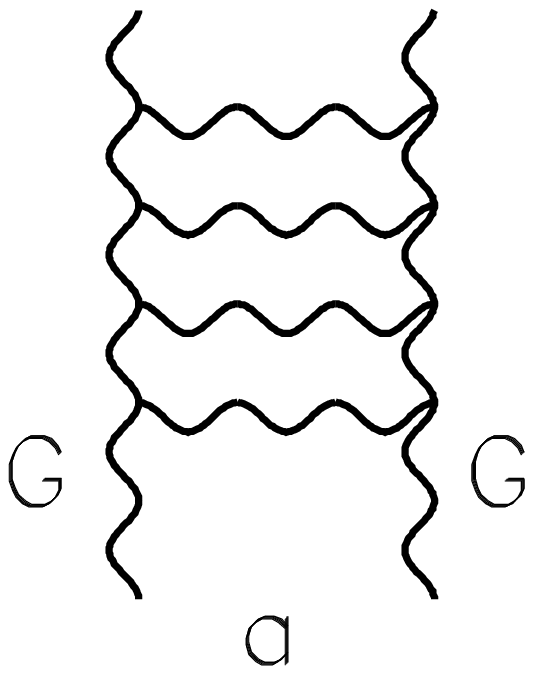,width=4cm}\hspace{2.5cm}
            \epsfig{file=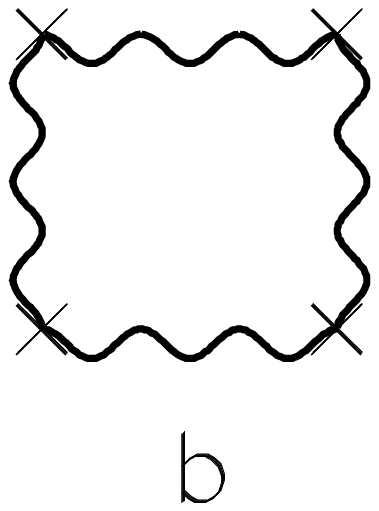,width=3cm}}
\vspace{1.5cm}
\centerline{\epsfig{file=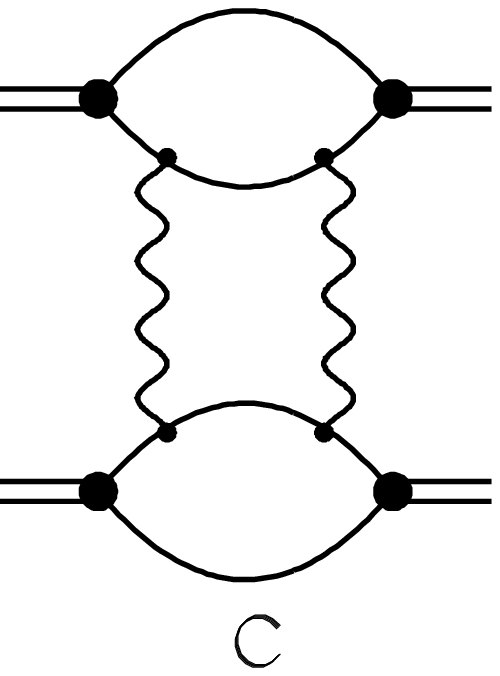,width=3cm}\hspace{0.5cm}
            \epsfig{file=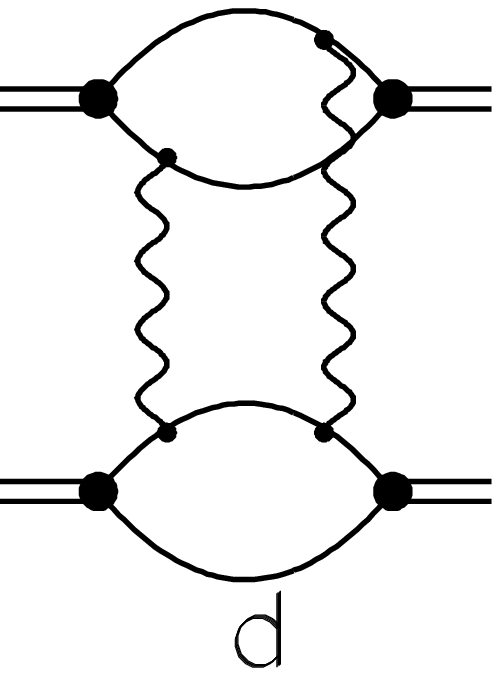,width=3cm}\hspace{0.5cm}
            \epsfig{file=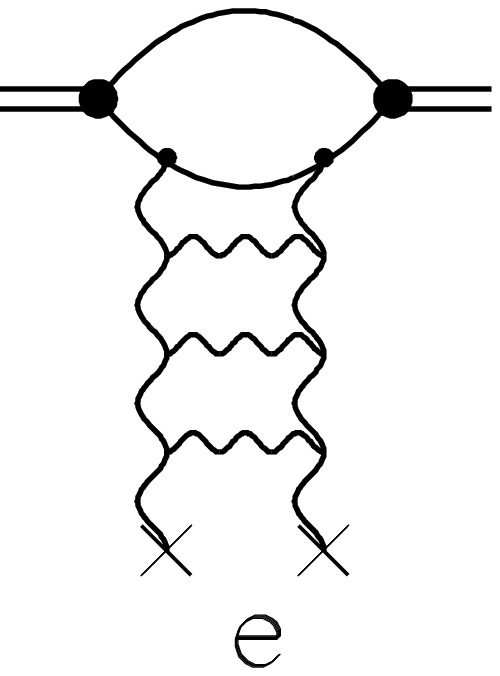,width=3cm}\hspace{0.5cm}
            \epsfig{file=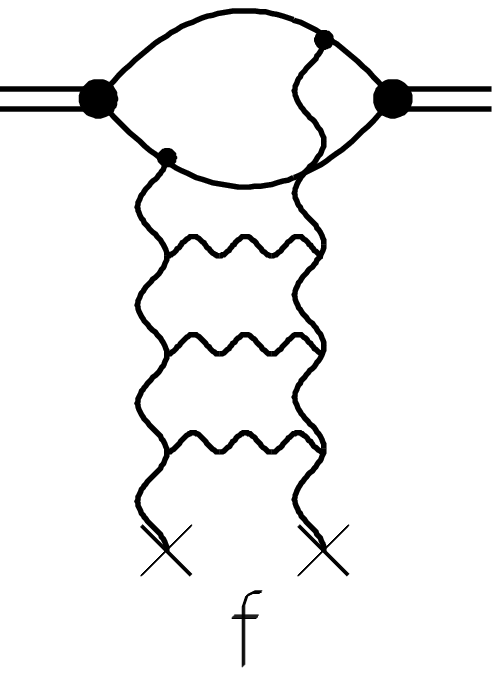,width=3cm}}
\vspace{0.5cm}
\caption{(a) Pomeron as gluon ladder; (b) gluon plaquet -- a
constructive element for the pomeron; (c)--(d) two-gluon exchange
diagrams for meson--meson scattering: impulse approximation (c) and
colour screening (d) diagrams; (e)--(f) pomerom--meson amplitude:
impulse approximation (e) and colour screening (f) diagrams.}
\end{figure}

\newpage
\begin{figure}
\centerline{\epsfig{file=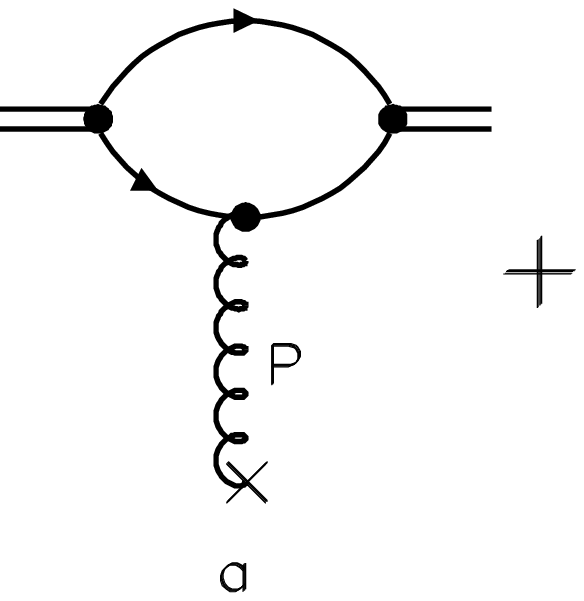,height=4cm}
            \epsfig{file=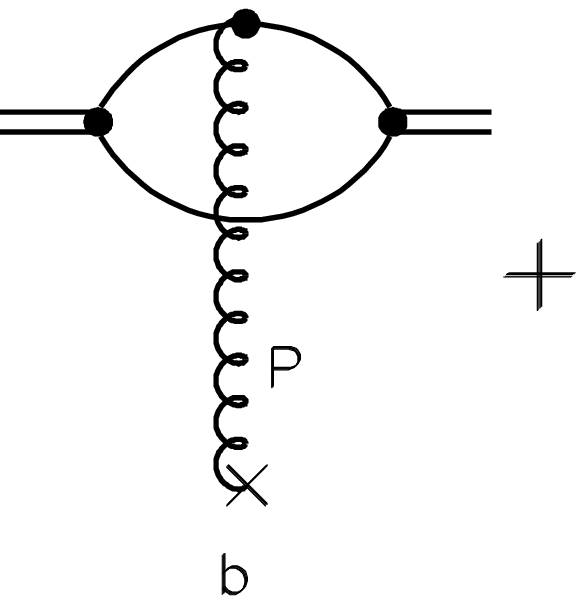,height=4cm}
            \epsfig{file=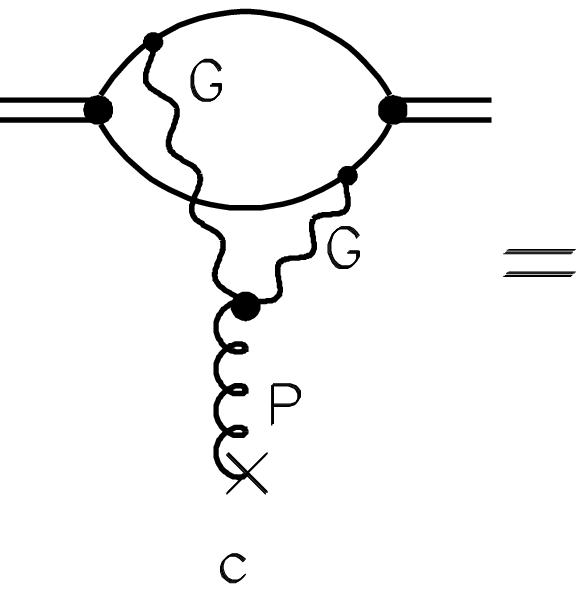,height=4cm}
            \epsfig{file=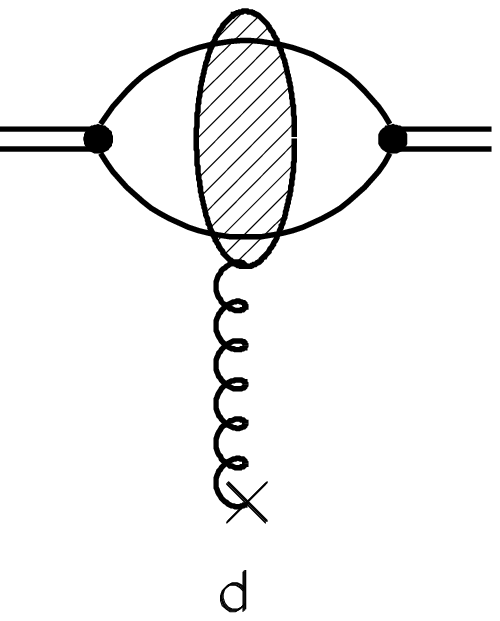,height=4cm}}
\vspace{1.5cm}
\centerline{\epsfig{file=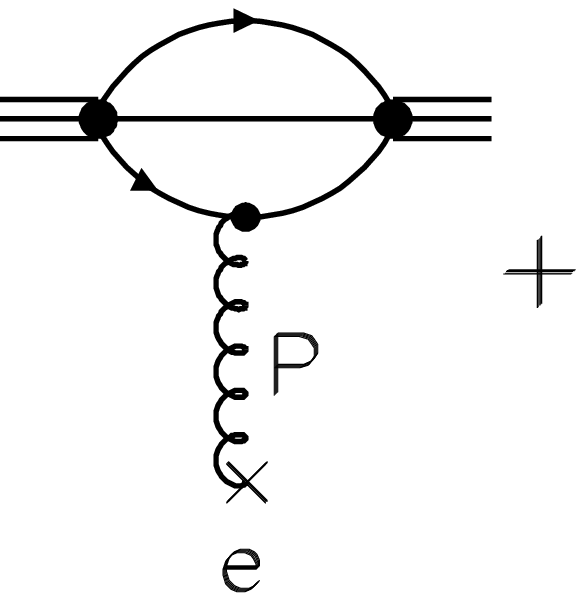,height=3cm}
            \epsfig{file=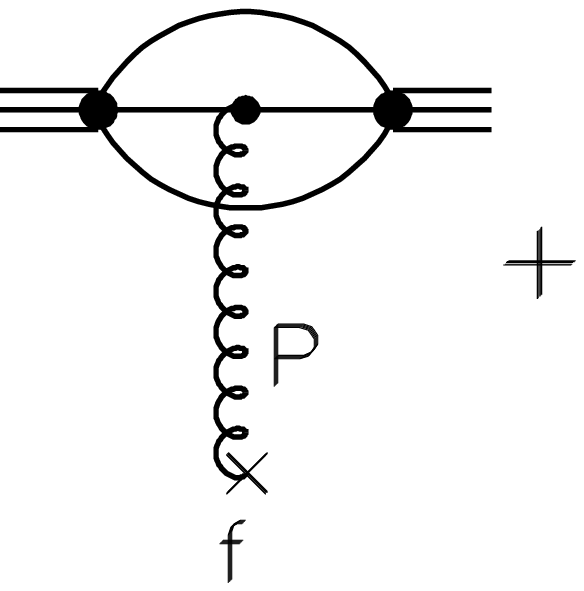,height=3cm}
            \epsfig{file=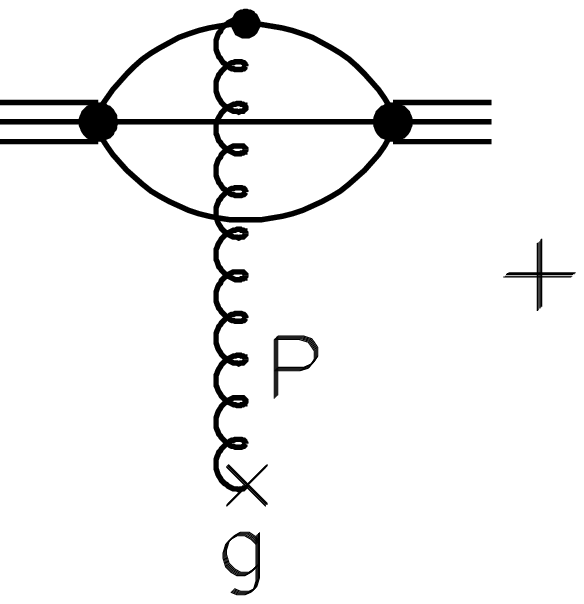,height=3cm}
            \epsfig{file=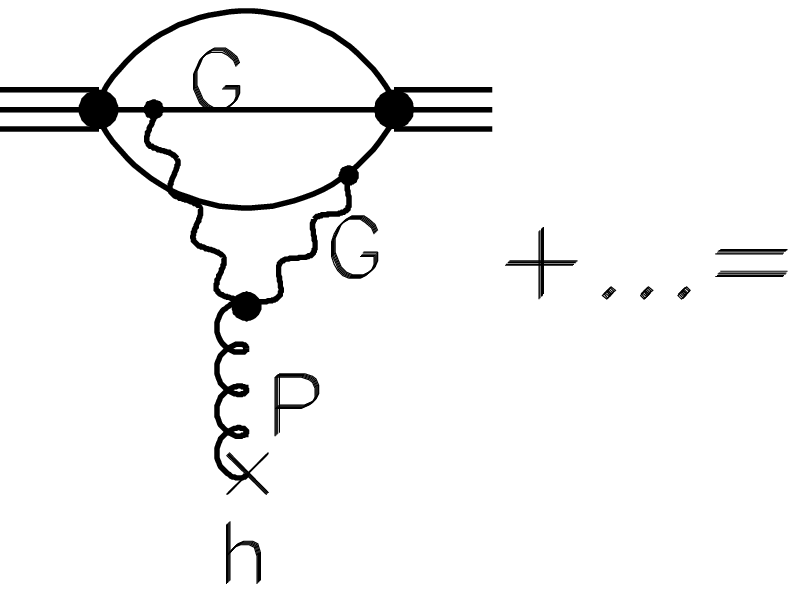,height=3cm}\hspace{-0.5cm}
            \epsfig{file=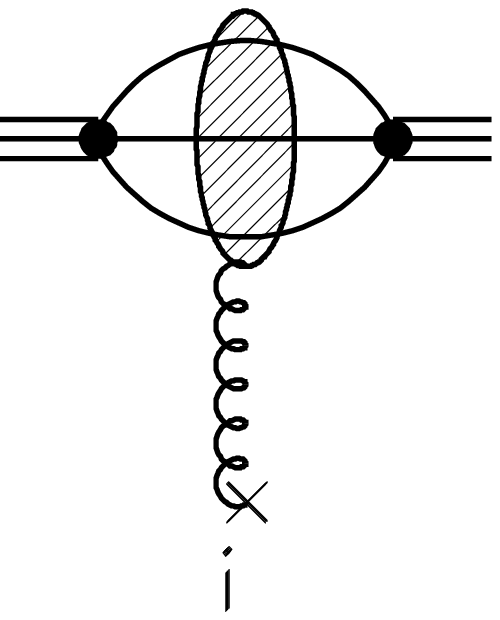,height=3cm}}
\vspace{0.5cm}
\caption{Diagrammatic representation of the pomeron--meson (a)--(d) and
pomeron--proton (e)--(i) amplitudes, with {\bf P} being a pomeron and
{\bf G} reggeized gluon. Three--reggeon diagrams {\bf PGG} provide
the colour screening.}
\end{figure}

\newpage
\begin{figure}
\centerline{\epsfig{file=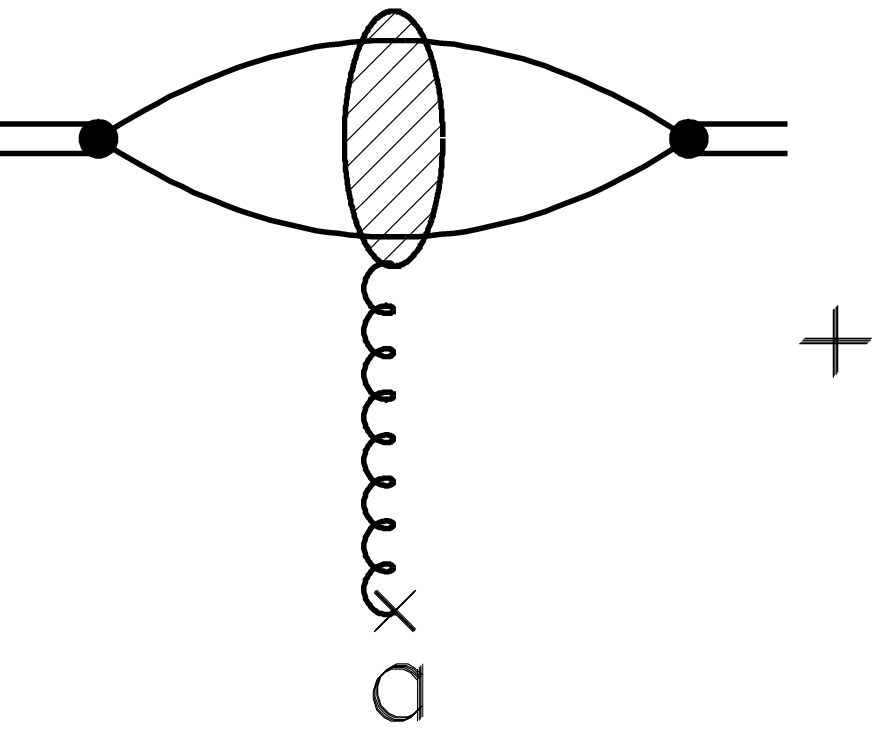,height=3.5cm}\hspace{0.5cm}
            \epsfig{file=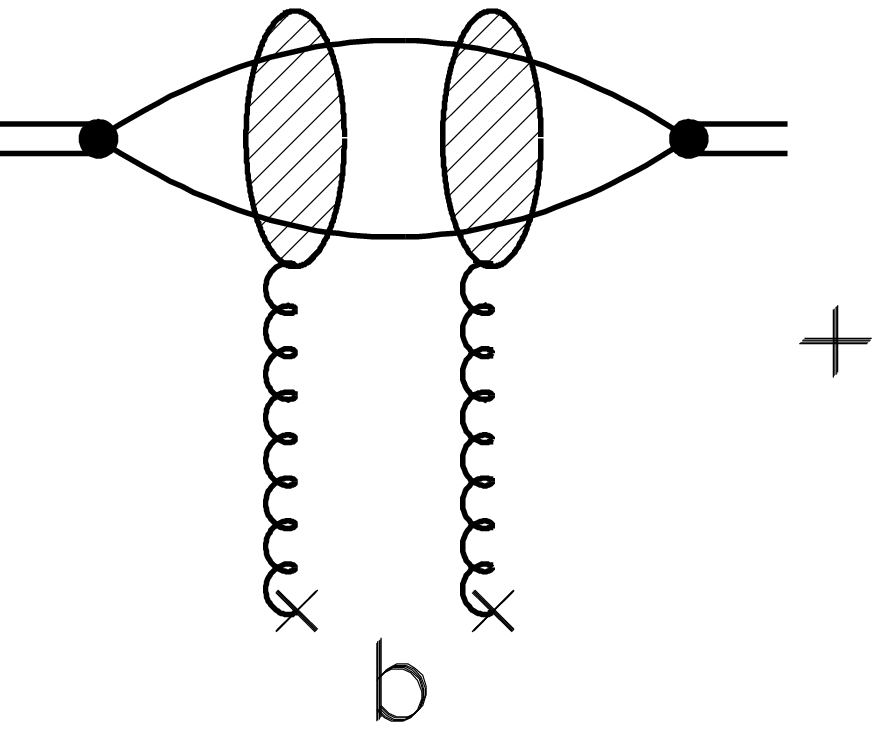,height=3.5cm}\hspace{0.5cm}
            \epsfig{file=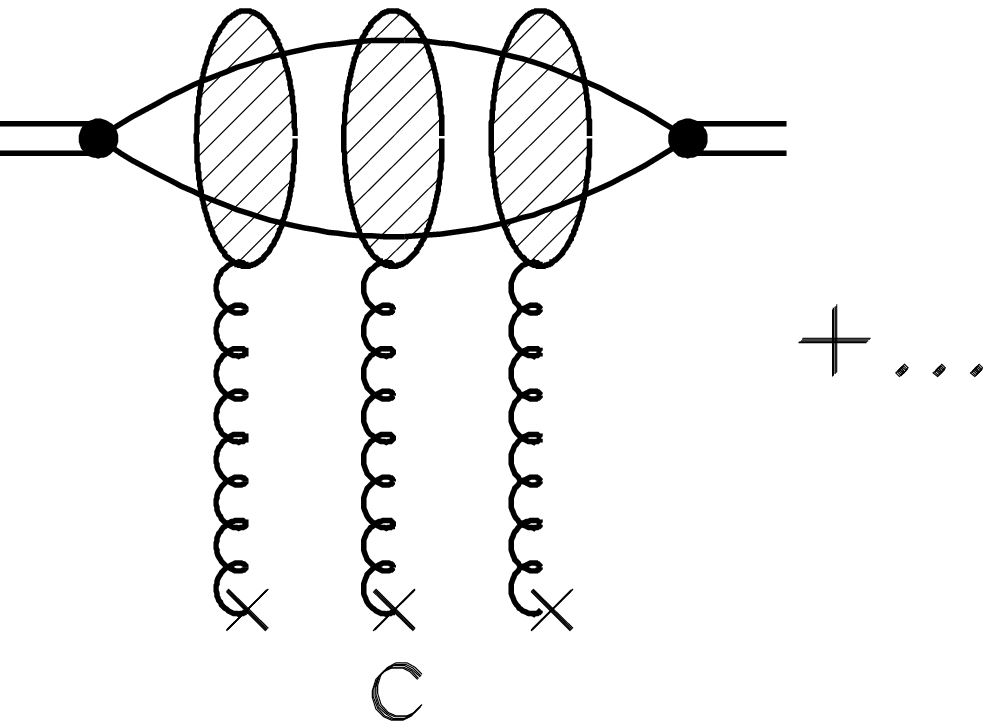,height=3.5cm}}
\vspace{0.5cm}
\caption{(a)--(c) Examples of multi-pomeron exchange diagrams,
colour screening term included; the shaded block is that of Fig. 2d.}
\end{figure}

\newpage
\begin{figure}
\centerline{\epsfig{file=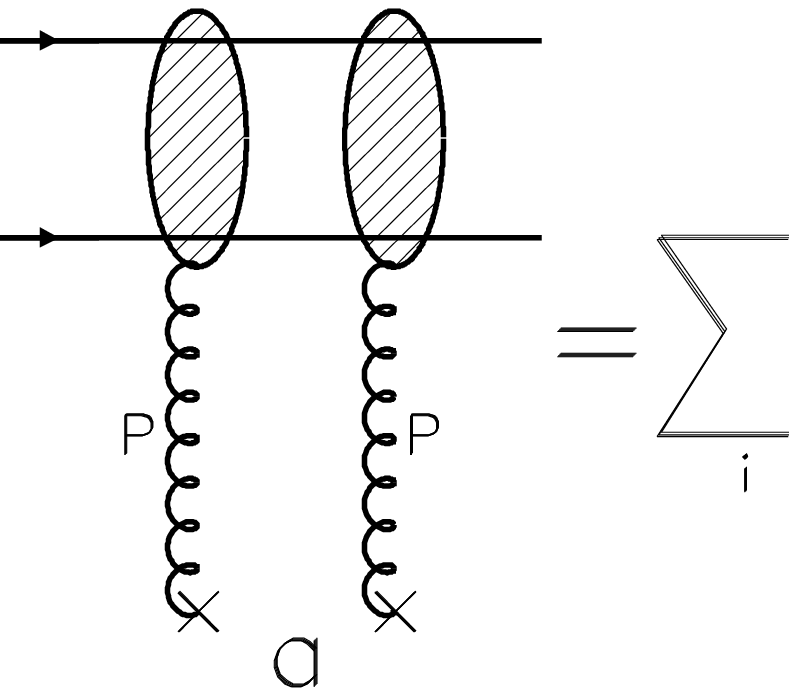,height=4cm}\hspace{0.3cm}
            \epsfig{file=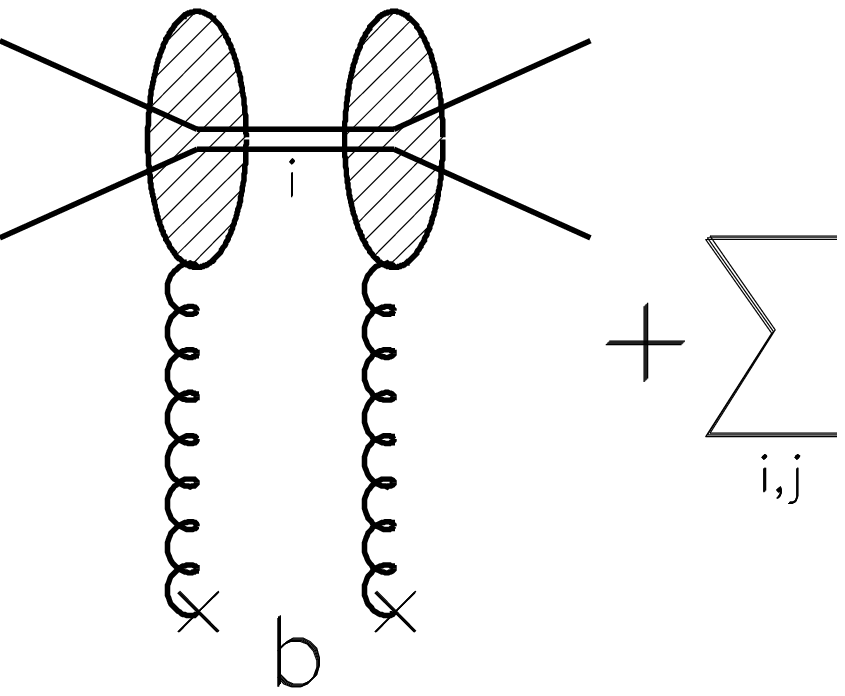,height=4cm}\hspace{0.3cm}
            \epsfig{file=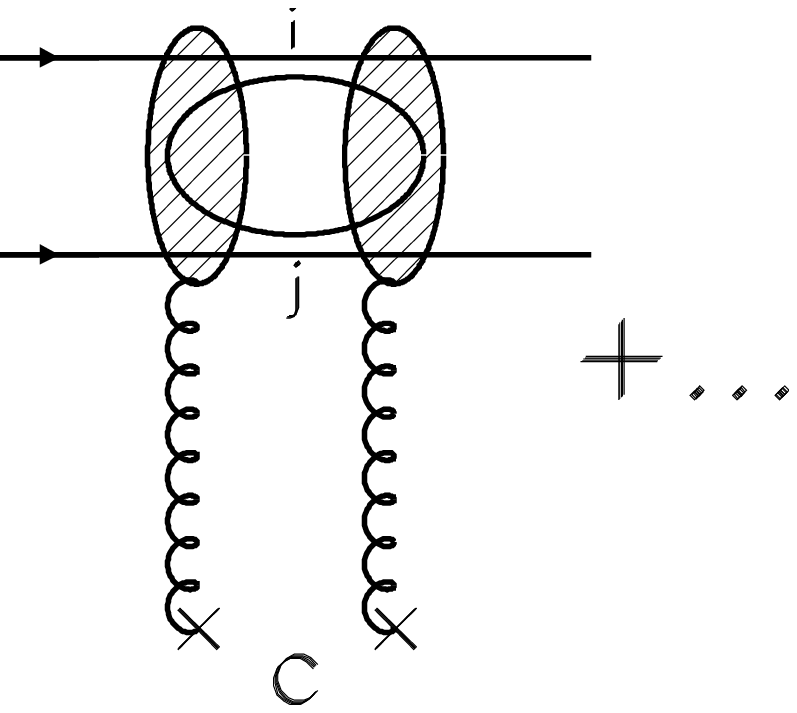,height=4cm}}
\vspace{1.5cm}
\centerline{\epsfig{file=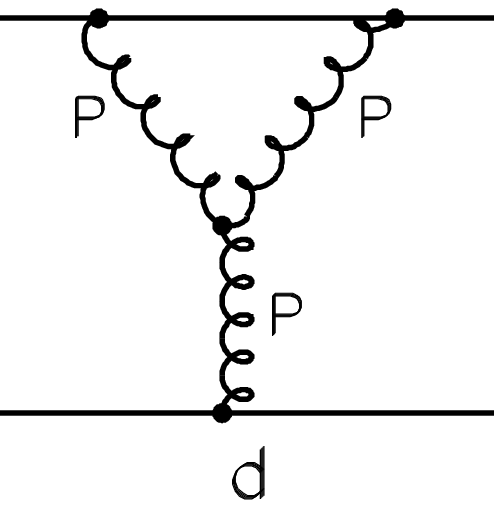,height=4cm}\hspace{2cm}
            \epsfig{file=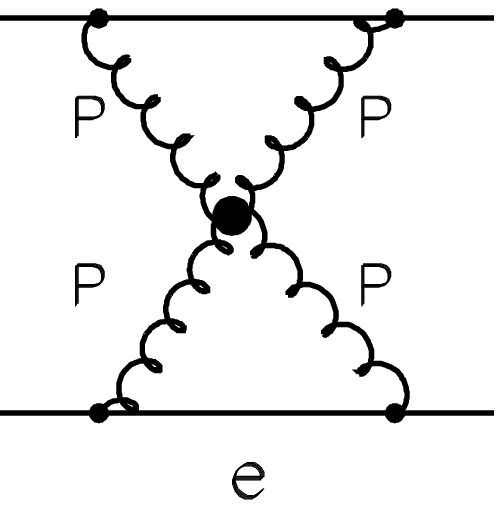,height=4cm}}
\vspace{0.5cm}
\caption{(a)--(c) Double rescattering diagrams: the intermediate
$q\bar q$ state is equivalent to the sum of all possible hadron states
(b), (c), etc.; (d) the three-pomeron ({\bf PPP}) $t$-channel exchange;
(e) the four-pomeron exchange amplitude {\bf PPPP}.}
\end{figure}

\newpage
\begin{figure}
\centerline{\epsfig{file=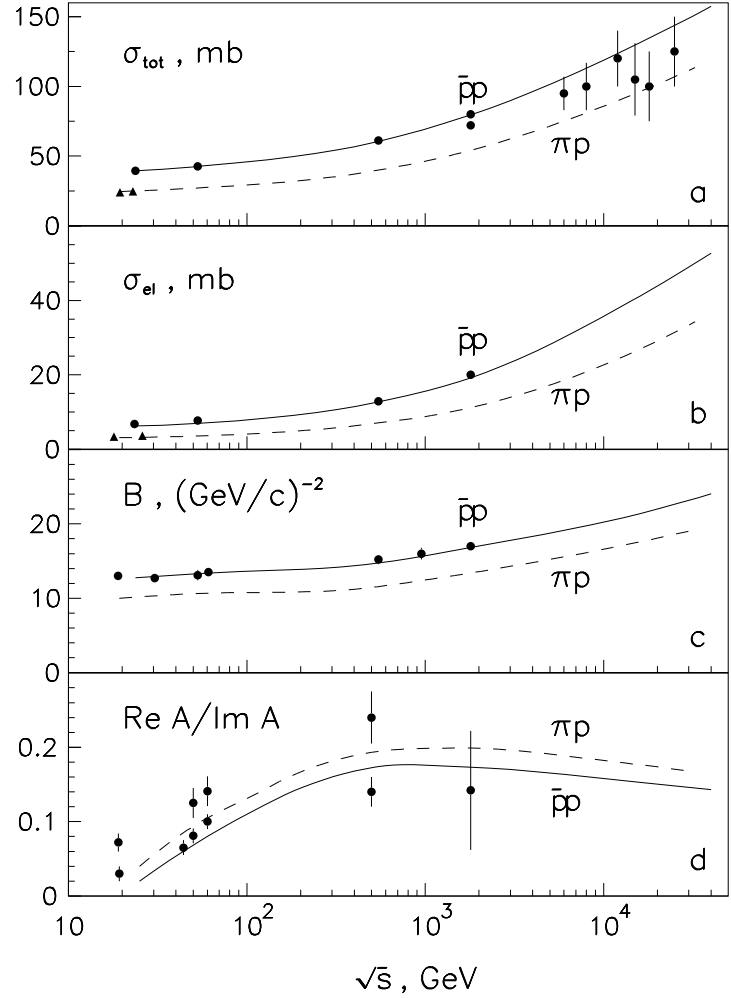,width=10cm}}
\vspace{0.5cm}
\caption{Description of experimental data in the energy range
$\sqrt{s}=20 - 10^5$ GeV for
(a) total and (b) elastic $p\bar p(pp)$ and $\pi p$ cross sections;
(c)  diffraction cone slope,
and (d) the ratio real/imaginary parts of the amplitude.}
\end{figure}

\newpage
\begin{figure}
\centerline{\epsfig{file=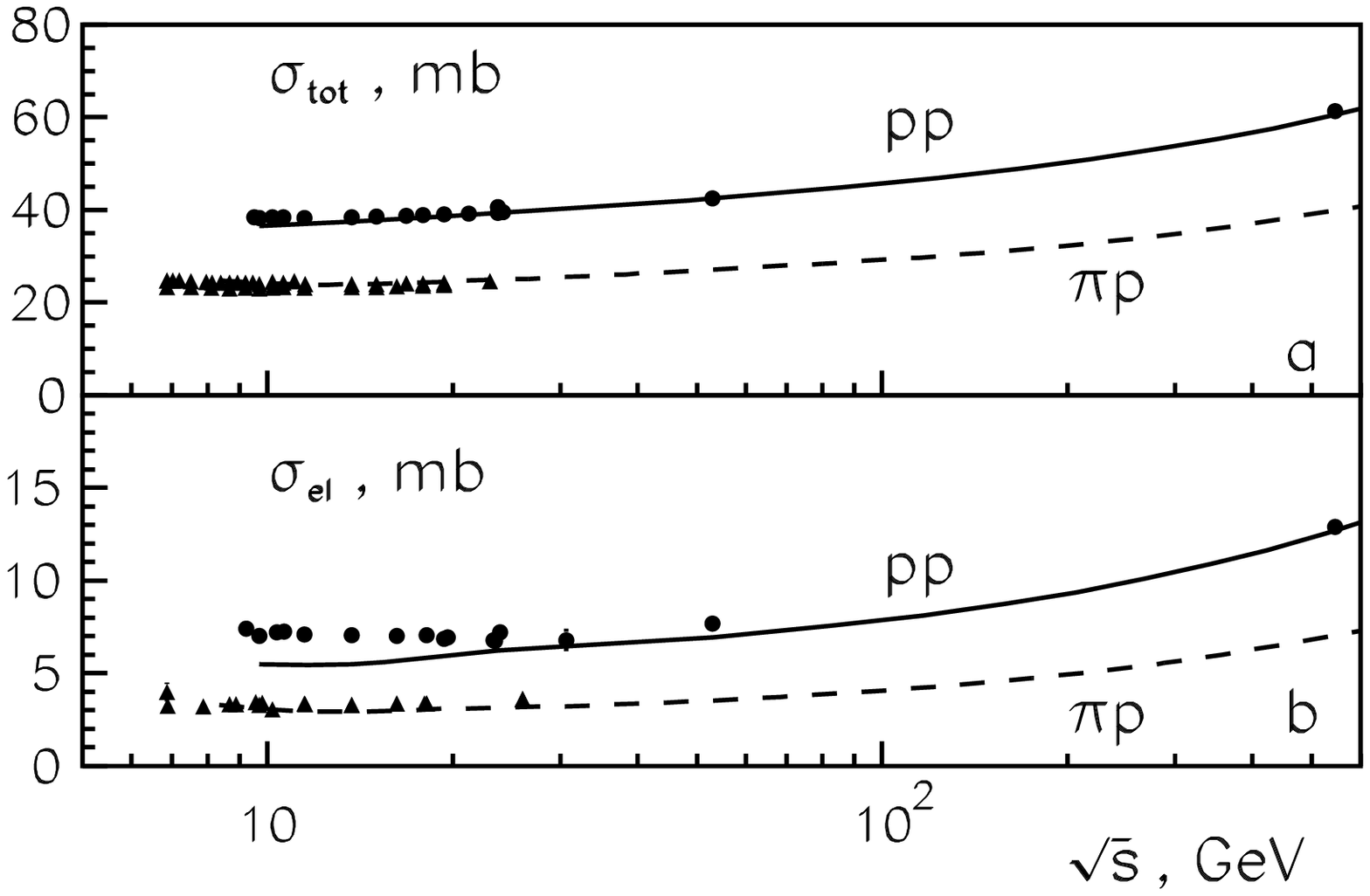,width=10cm}}
\vspace{0.5cm}
\centerline{\epsfig{file=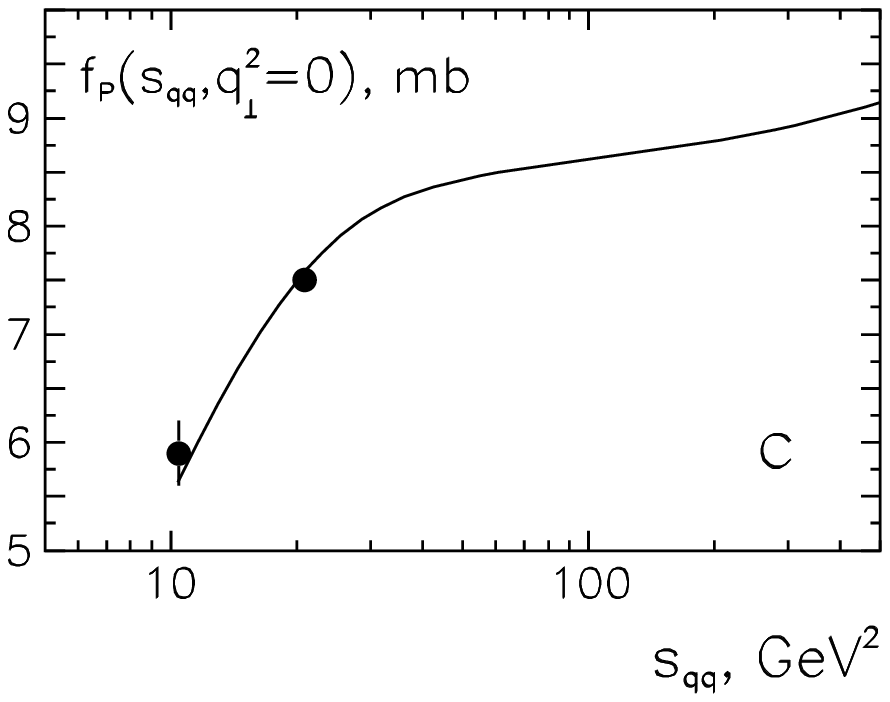,width=6.5cm}
            \epsfig{file=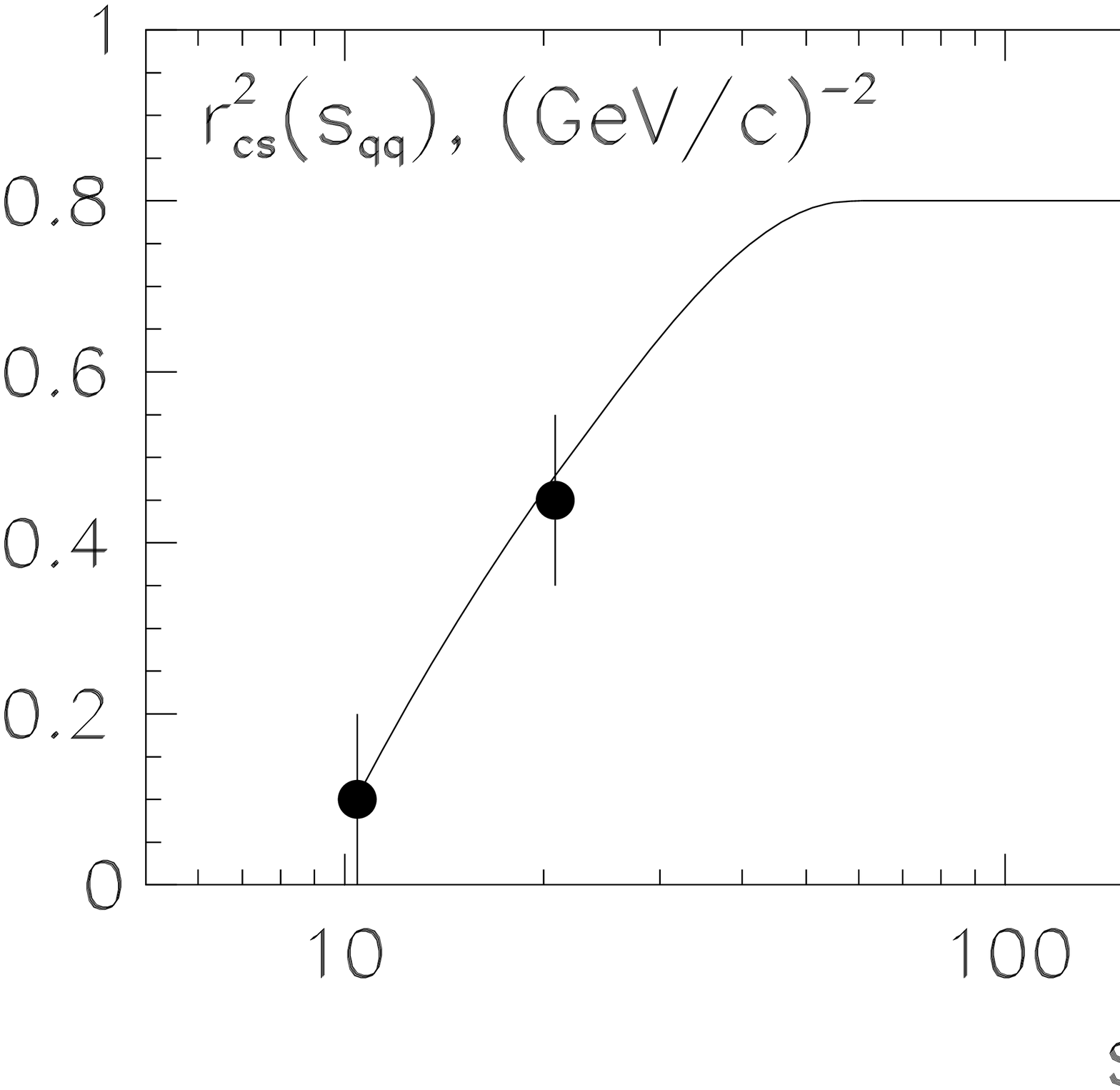,width=6.5cm}}
\vspace{0.5cm}
\centerline{\epsfig{file=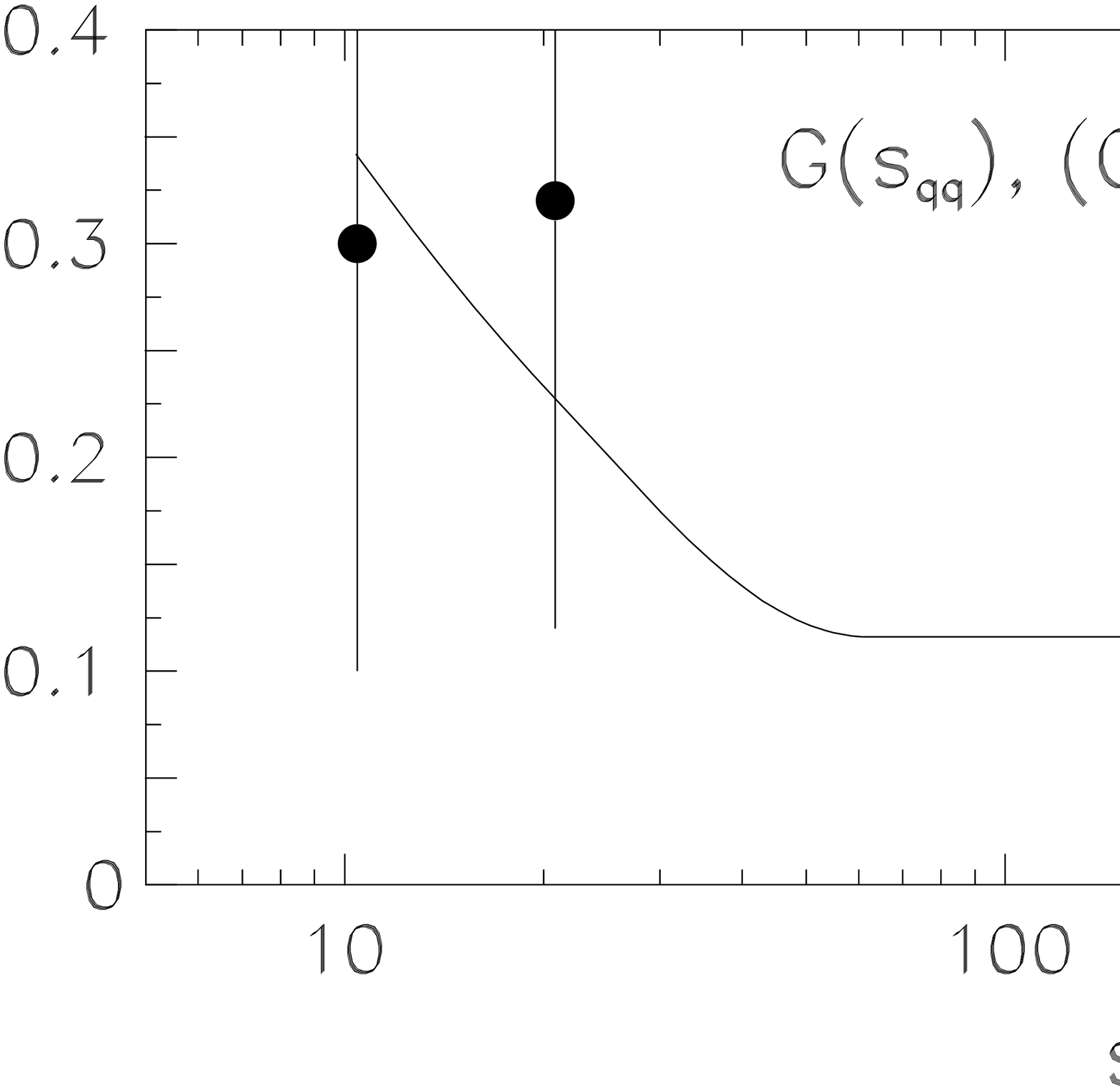,width=6.5cm}}
\vspace{0.5cm}
\caption{Description of experimental data in the energy range
$\sqrt{s}=5 - 25$ GeV for
(a) total and (b) elastic $p\bar p(pp)$ and $\pi p$ cross sections;
(c)--(e) primary pomeron parameters
as functions of the quark--quark energy
squared, $s_{qq}$: (c) quark--pomeron coupling squared,
$f_P(s_{qq},q^2_\perp=0)$,
(d) colour
screening radius squared, $r^2_{cs}$, and (e) the pomeron slope $G$. The
error bars show uncertainties of the parameter definition.}
\end{figure}

\begin{figure}
\centerline{\epsfig{file=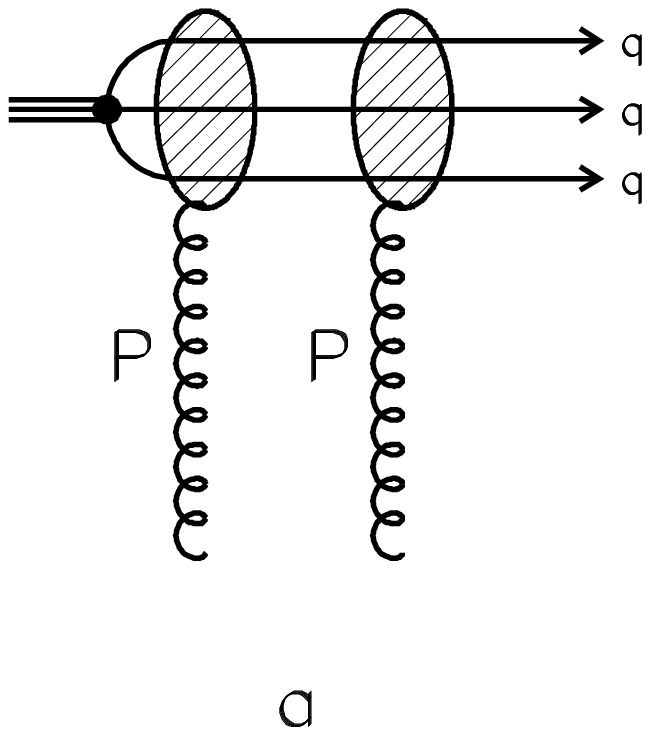,height=5cm}
            \epsfig{file=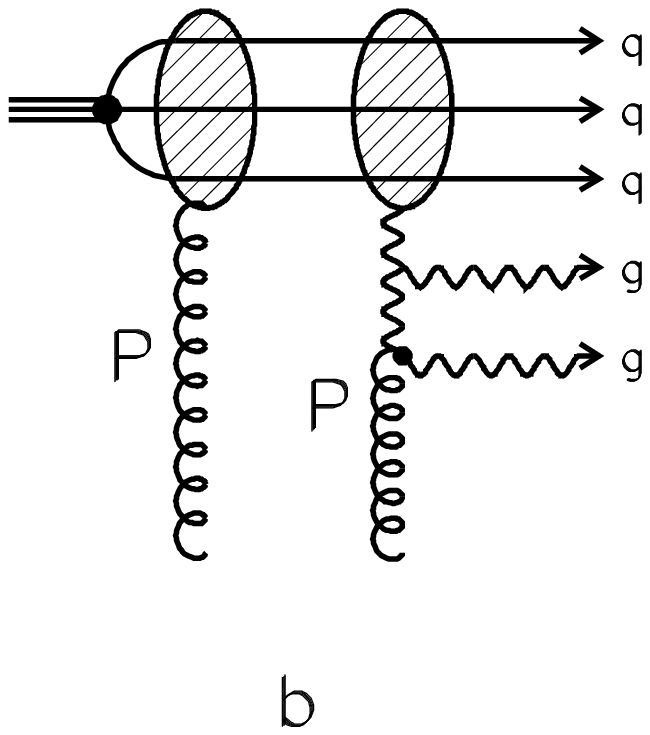,height=5cm}\hspace{0.5cm}
            \epsfig{file=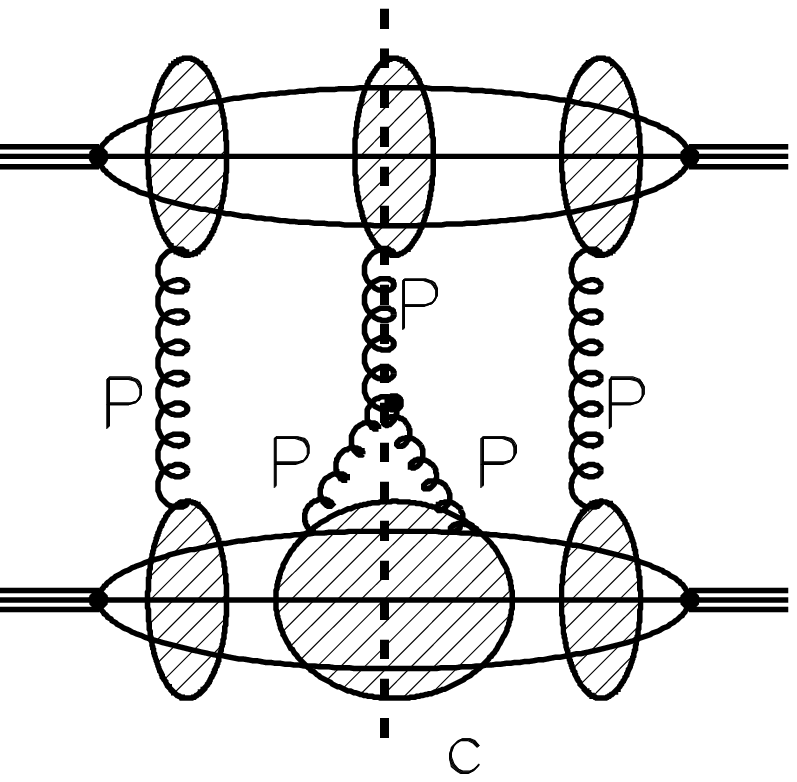,height=5cm}}
\vspace{1cm}
\centerline{\epsfig{file=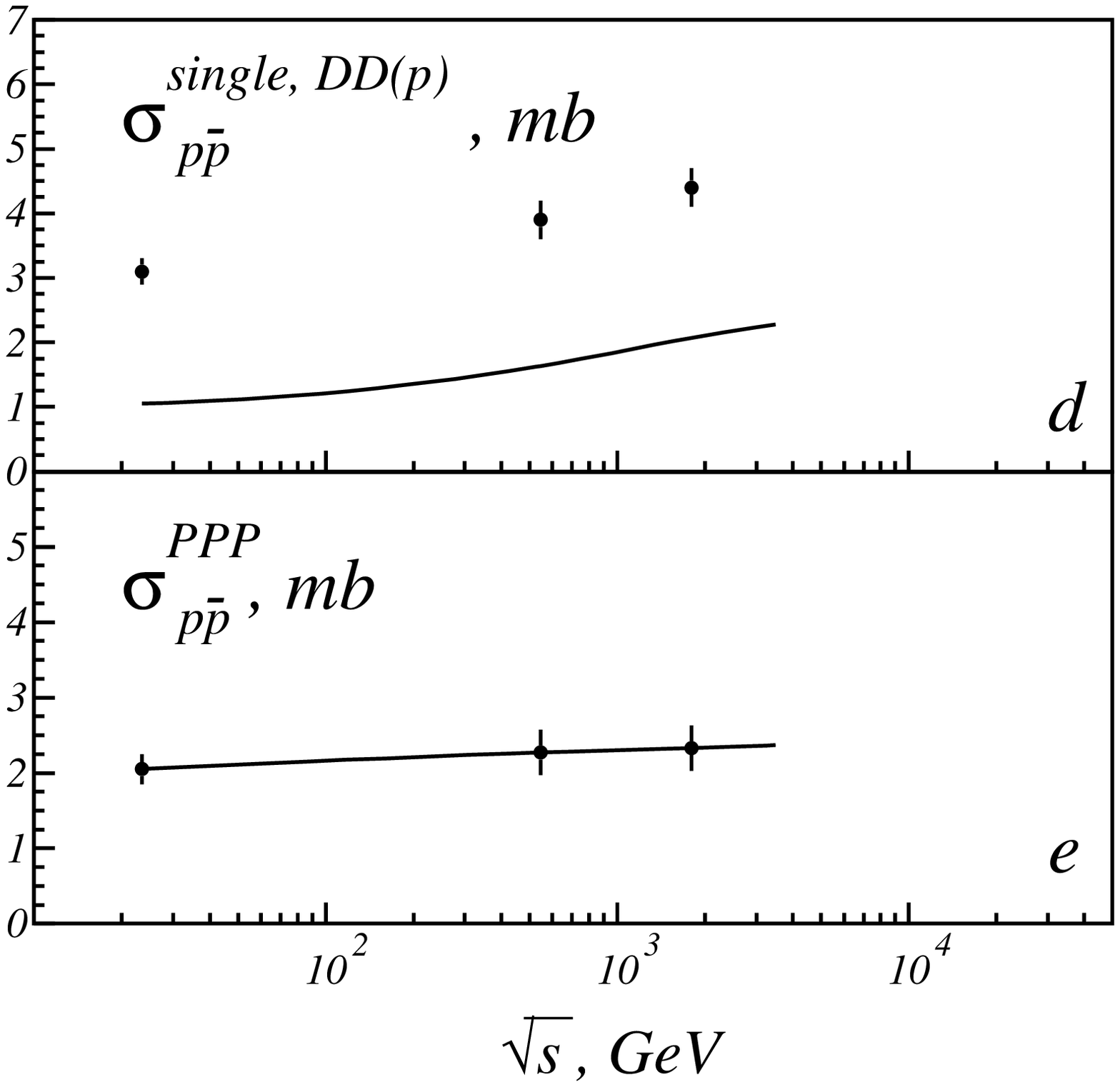,width=10cm}}
\caption{(a)-(b) Diagrams describing single diffraction dissociation:
dissociation of a proton (a) and partly dissociating pomeron (b);
(c) cutting of {\bf PPP}-diagramm that gives the cross section with
partly dissociating pomeron.
(d) Experimental data for single diffractive dissociation of a proton,
solid curve stands for the result of the calculation according to
(20); (e) the estimated contribution for the dissociation of pomeron
(three--pomeron diagram, Fig. 7a).}
\end{figure}

\begin{figure}
\centerline{\epsfig{file=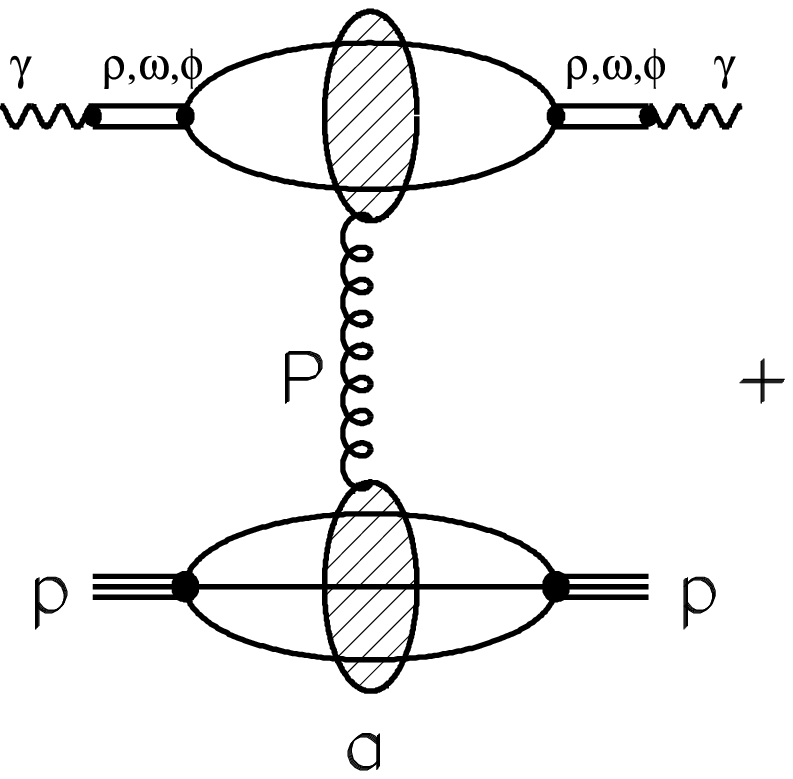,width=5cm}\hspace{0.2cm}
            \epsfig{file=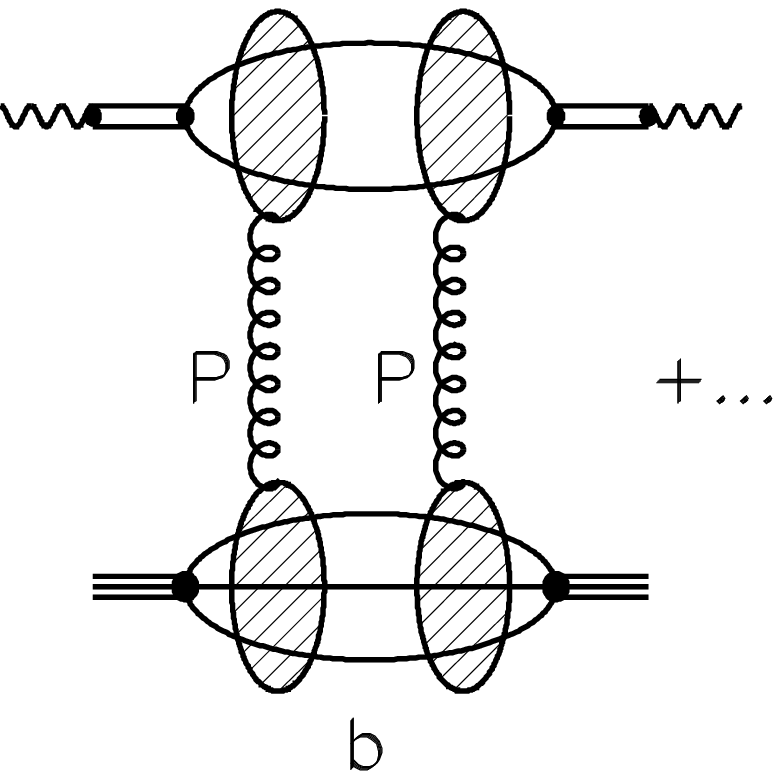,width=5cm}}
\vspace{1cm}
\centerline{\epsfig{file=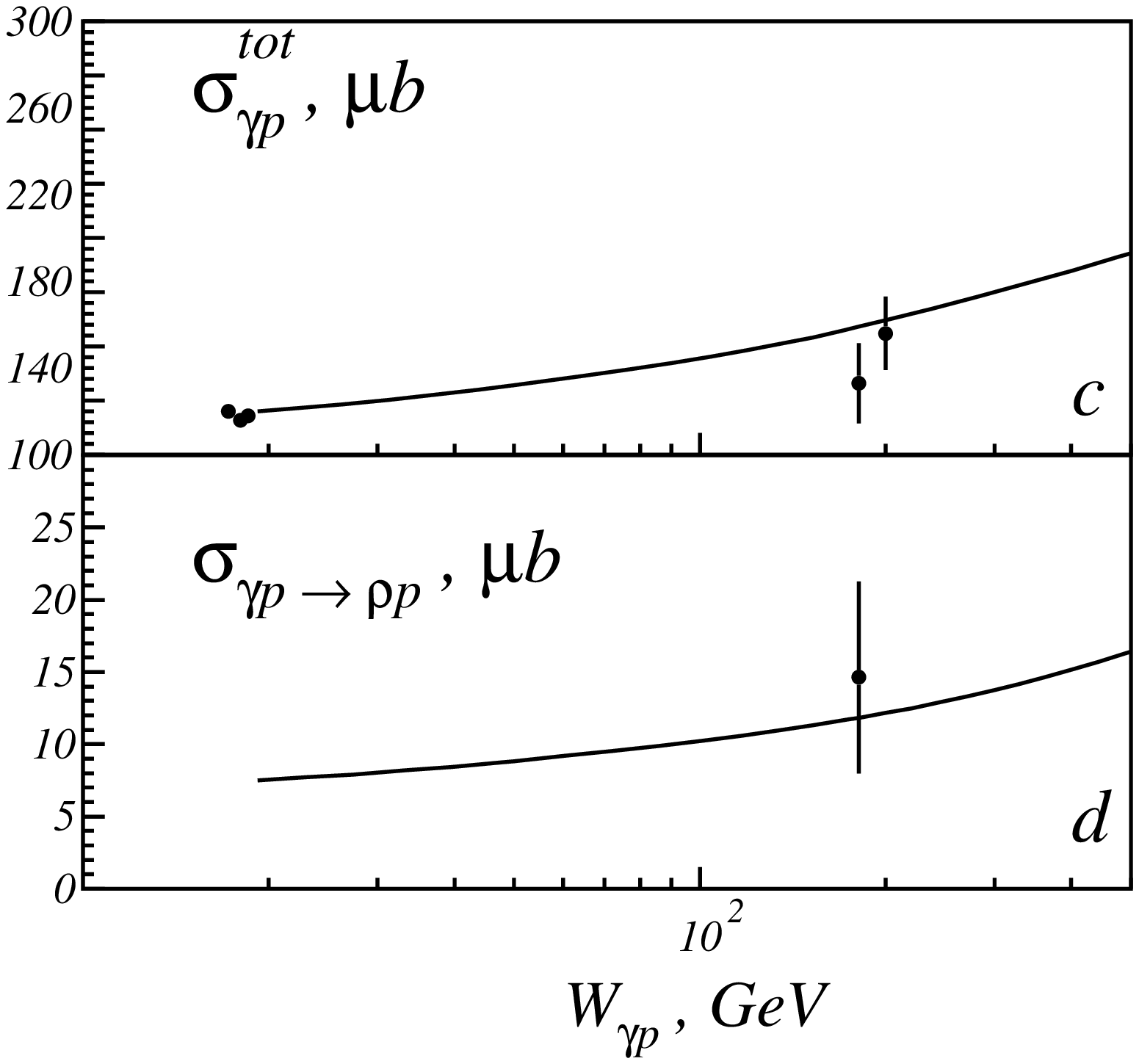,width=10cm}}
\caption{Photon-proton collisions: (a)--(b) diagrammatic
representation of the $\gamma p$ scattering
amplitude; (c) description of the data on total cross section
$\gamma p$, and (d) production of vector mesons  $\gamma p \to \rho
p$.}
\end{figure}

\begin{figure}
\centerline{\epsfig{file=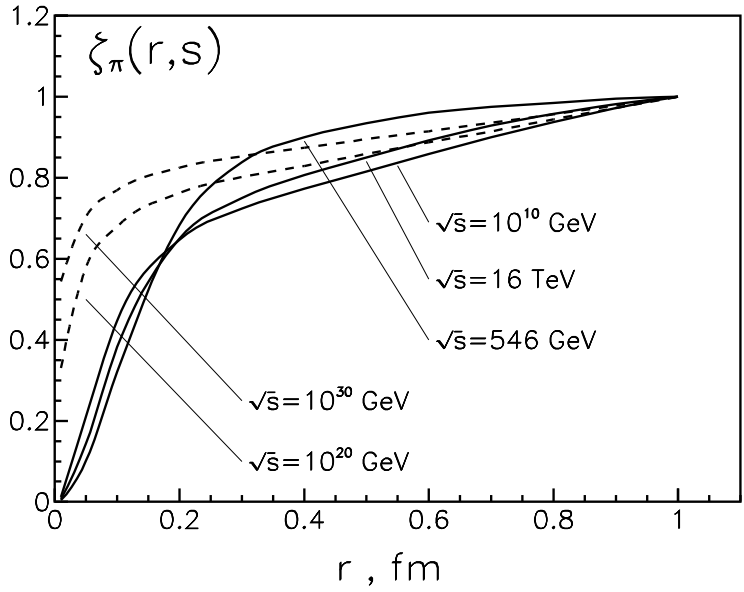,width=10cm}}
\caption{The pion profile function $\zeta_\pi (r,s)$
depending on the interquark distance
$r=|\vec r_{1\perp}-\vec r_{2\perp}|$ at different energies $\sqrt{s}$.}
\end{figure}

\newpage
\begin{figure}
\centerline{\epsfig{file=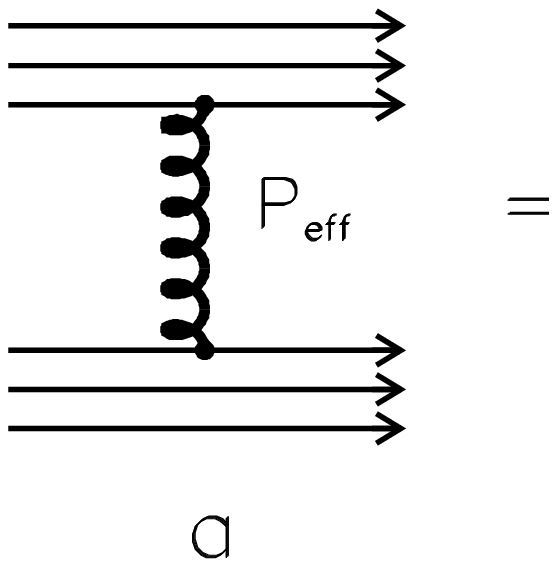,height=4.3cm}\hspace{0.5cm}
            \epsfig{file=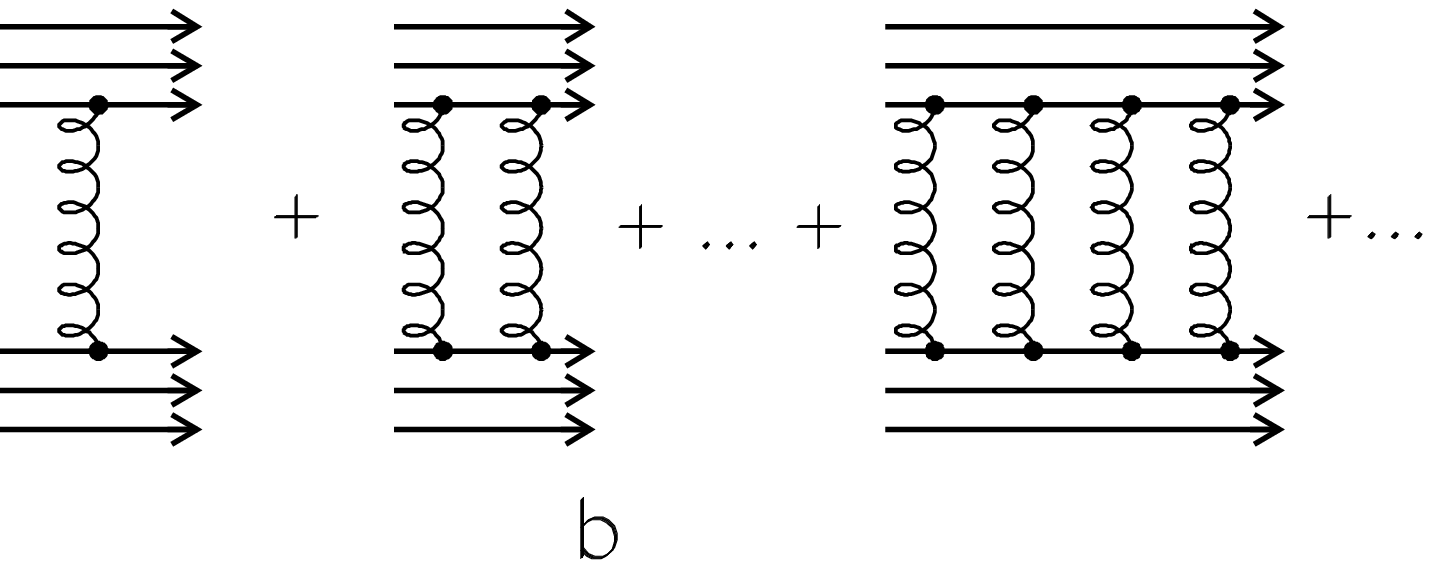,height=4cm}}
\vspace{1.5cm}
\centerline{\epsfig{file=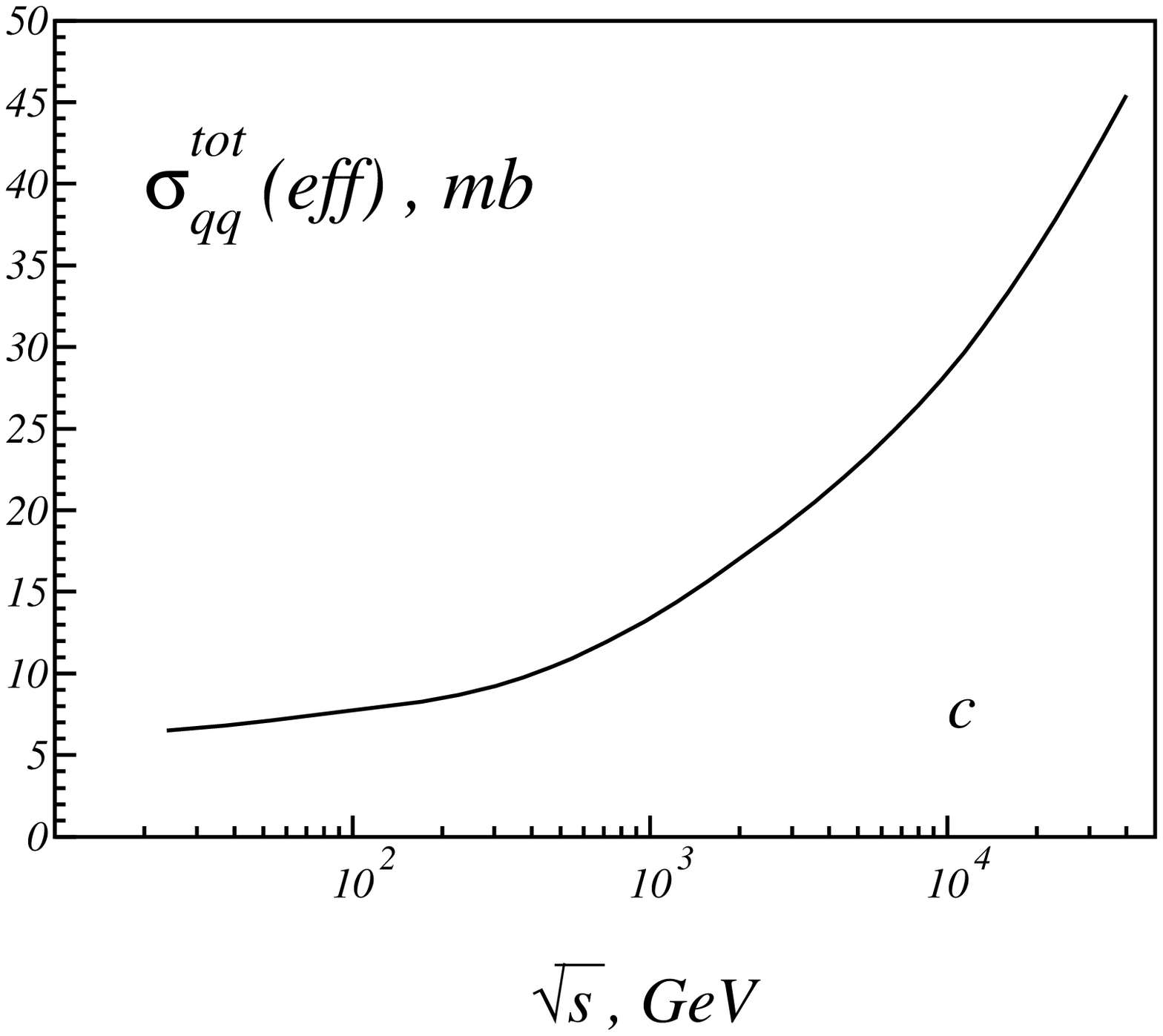,width=9cm}}
\caption{(a)--(b) Effective pomeron {\bf P$_{eff}$} as a sum of the
exchanges of primary pomerons and (c) the energy
dependence of the quark--quark amplitude
$f^{(P_{eff})}_{qq}=\sigma^{tot}_{qq}(eff)$  due to the exchange of
the effective pomeron.} \end{figure}

\newpage
\begin{figure}
\centerline{\epsfig{file=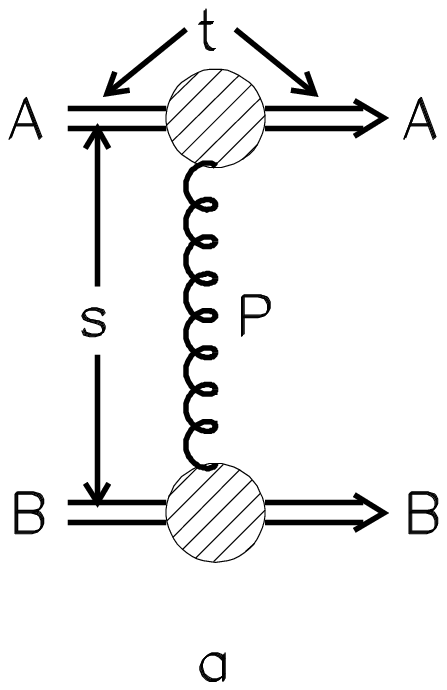,height=4cm}\hspace{0.5cm}
            \epsfig{file=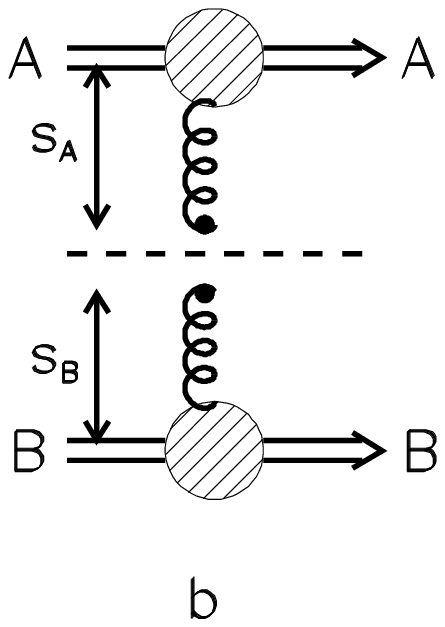,height=3.8cm}}
\caption{(a) Meson--meson scattering amplitude;
(b) the pomeron cut enables to treat the
pomeron--meson amplitudes separately.}
\end{figure}

\newpage
\begin{figure}
\centerline{\epsfig{file=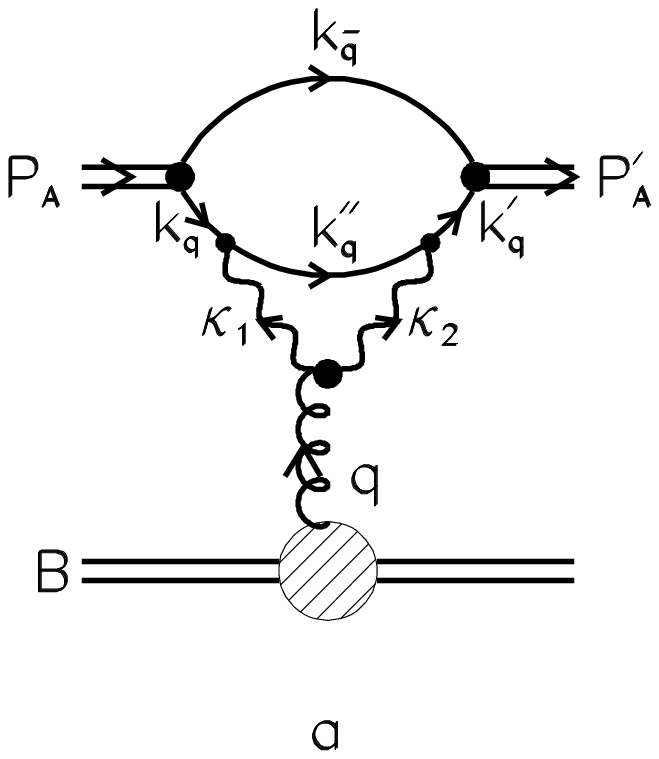,height=5cm}\hspace{1cm}
            \epsfig{file=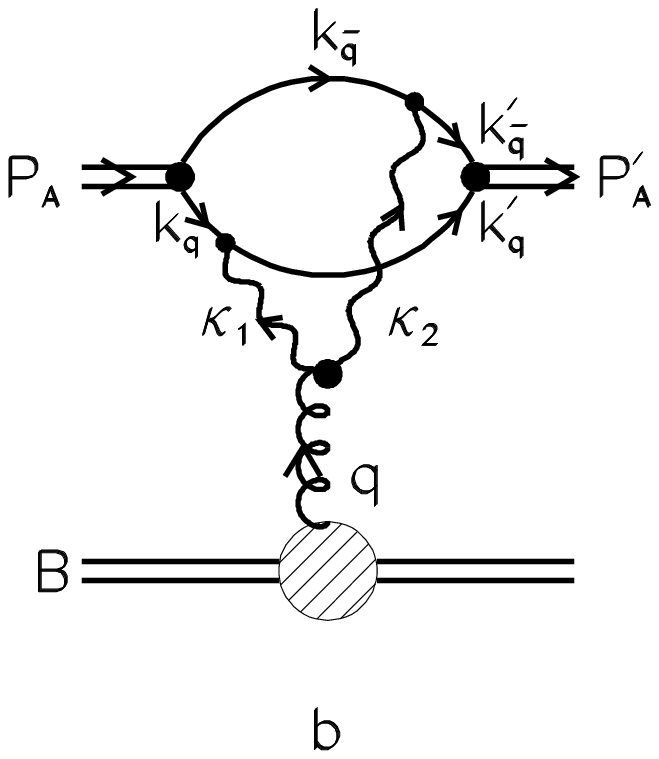,height=5cm}}
\vspace{0.5cm}
\centerline{\epsfig{file=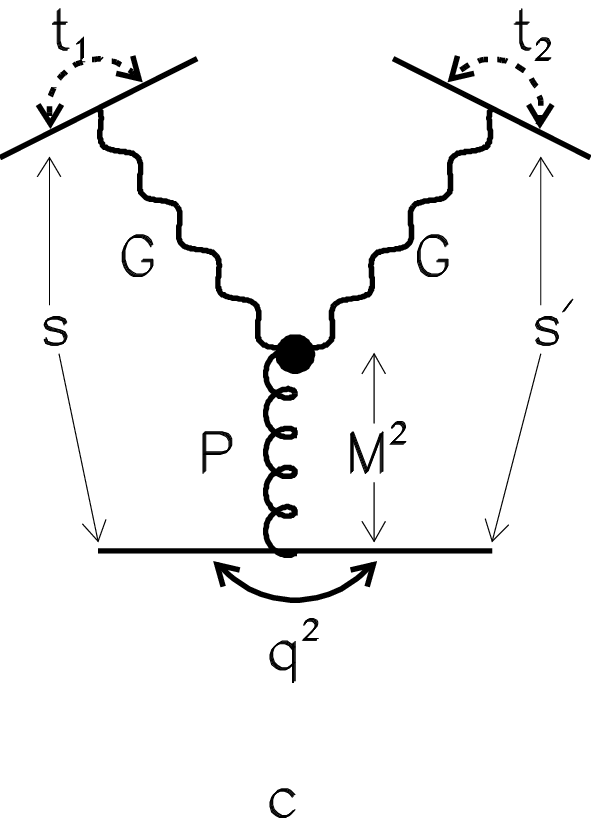,height=5cm}\hspace{1cm}
            \epsfig{file=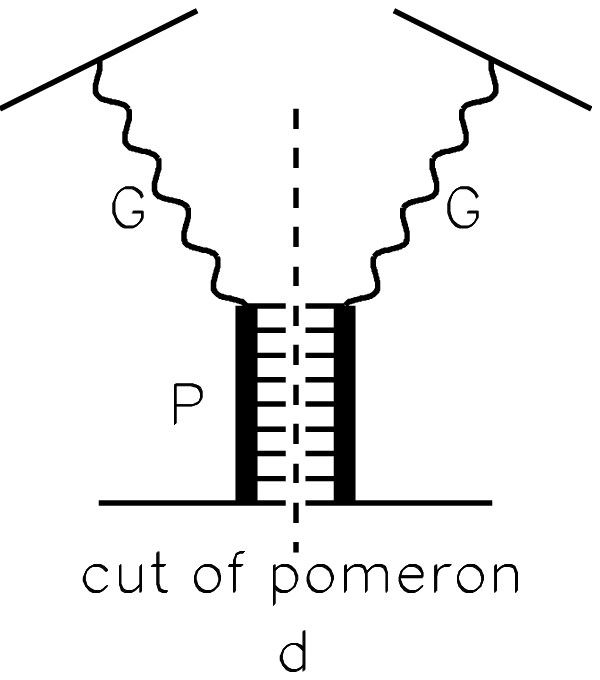,height=5cm}}
\vspace{0.5cm}
\caption{The {\bf PGG} amplitude for coupling to (a) a single quark,
(b) quark and antiquark; (c) the used notations for the {\bf PGG}
diagram; (d) cutting along the pomeron line represents the
discontinuity of the amplitude, $disc_{M^2}\,A_{{\bf PGG}}$.}
\end{figure}

\end{document}